\documentclass[prd,superscriptaddress,floatfix,preprintnumbers,altaffilletter,twocolumn,nofootinbib]{revtex4}
\usepackage{graphicx,url,amssymb,amsmath,longtable,rotating,color,units,
wasysym,subfigure,epsfig,cancel,placeins}
\usepackage[mathscr]{euscript}
\usepackage{color}
\usepackage{verbatim}
\usepackage{dcolumn}
\usepackage{bm}
\usepackage{epstopdf}

\newcommand*\dif{\mathop{}\!\mathrm{d}}

\begin{document}

\title{A Mock Data and Science Challenge for Detecting an Astrophysical Stochastic Gravitational-Wave Background with Advanced LIGO and Advanced Virgo}
\author{Duncan Meacher}
\email{Duncan.Meacher@ligo.org}
\affiliation{UMR ARTEMIS, CNRS, University of Nice Sophia-Antipolis, Observatoire de la C$\hat{o}$te d'Azur, BP 4229, 06304, Nice Cedex 4, France}

\author{Michael Coughlin}
\affiliation{Department of Physics, Harvard University, Cambridge, MA 02138, USA}

\author{Sean Morris}
\affiliation{Department of Physics and Astronomy and Center for Gravitational-Wave Astronomy, University of Texas at Brownsville, Brownsville, Texas 78520, USA}

\author{Tania Regimbau}
\affiliation{UMR ARTEMIS, CNRS, University of Nice Sophia-Antipolis, Observatoire de la C$\hat{o}$te d'Azur, BP 4229, 06304, Nice Cedex 4, France}

\author{Nelson Christensen}
\affiliation{Physics and Astronomy, Carleton College, Northfield, Minnesota 55057, USA}

\author{Shivaraj Kandhasamy}
\affiliation{Physics and Astronomy, University of Mississippi, University, MS 38677-1848, USA}

\author{Vuk Mandic}
\affiliation{School of Physics and Astronomy, University of Minnesota, Minneapolis, Minnesota 55455, USA}

\author{Joseph D. Romano}
\affiliation{Department of Physics and Astronomy and Center for Gravitational-Wave Astronomy, University of Texas at Brownsville, Brownsville, Texas 78520, USA}

\author{Eric Thrane}
\affiliation{School of Physics and Astronomy, Monash University, Clayton, Victoria 3800, Australia}

\date{\today}

\begin{abstract}
The purpose of this mock data and science challenge is to prepare the data analysis and science interpretation for the second generation of gravitational-wave experiments Advanced LIGO-Virgo in the search for a stochastic gravitational-wave background signal of astrophysical origin.
Here we present a series of signal and data challenges, with increasing complexity, whose aim is to test the ability of current data analysis pipelines at detecting an astrophysically produced gravitational-wave background, test parameter estimation methods and interpret the results.
We introduce the production of these mock data sets that includes a realistic observing scenario data set where we account for different sensitivities of the advanced detectors as they are continuously upgraded toward their design sensitivity.
After analysing these with the standard isotropic cross-correlation pipeline we find that we are able to recover the injected gravitational-wave background energy density to within $2\sigma$ for all of the data sets and present the results from the parameter estimation.
The results from this mock data and science challenge show that advanced LIGO and Virgo will be ready and able to make a detection of an astrophysical gravitational-wave background within a few years of operations of the advanced detectors, given a high enough rate of compact binary coalescing events.
\end{abstract}

\maketitle

\section{Introduction}
\label{sec:introduction}

According to various cosmological and astrophysical scenarios, we are bathed in a stochastic gravitational wave background (SGWB).
Proposed theoretical cosmological models include the amplification of vacuum fluctuations during inflation\cite{spjetp.40.409.75,jetpl.30.682.79,prd.48.3513.93}, pre Big Bang models~\cite{app.1.317.93,prd.55.3330.97,prd.82.083518.10}, cosmic (super)strings~\cite{prd.71.063510.05,prl.98.111101.07,prd.81.104028.10,prd.85.066001.12}, and phase transitions~\cite{prd.77.124015.08,prd.79.083519.09,jcap.12.024.09}.
In addition to the cosmological background~\cite{pr.331.283.00,jcap.6.27.12}, an astrophysical contribution~\cite{raap.11.369.11} is expected to result from the superposition of a large number of unresolved sources, such as core collapse supernovae to neutron stars or black holes~\cite{prd.72.084001.05,prd.73.104024.06,mnras.398.293.09,mnrasl.409.L132.10}, rotating neutron stars~\cite{aap.376.381.01,prd.86.104007.12}, including magnetars~\cite{aap.447.1.06,mnras.410.2123.10,mnras.411.2549.11,prd.87.042002.13}, phase transitions~\cite{grg.41.1389.09}, or initial instabilities in young neutron stars~\cite{mnras.303.258.99,mnras.351.1237.04,apj.729.59.11}, or compact binary mergers~\cite{apj.739.86.11,prd.84.084004.11,prd.84.124037.11,prd.85.104024.12,mnras.431.882.13}.

Many of these models are within reach of the next generation of gravitational-wave (GW) detectors such as Advanced LIGO Hanford (H) and Livingston (L)~\cite{cqg.32.074001.15}, which are expected to start collecting data in 2015, and Advanced Virgo (V)~\cite{cqg.32.024001.15}, which will begin collecting data in 2016.
These detectors are expected to have a final sensitivity 10 times better than that of the initial detectors~\cite{rpp.72.076901.09,aip.794.307.05}, which will be achieved over a period of several years of continued upgrades.
The detection of a cosmological background would provide very important constraints on the first instant of the Universe, up to the limits of the Planck era and the Big Bang, while the detection of an astrophysical background would provide crucial information on the physical properties of the respective astrophysical populations, the evolution of these objects with redshift, the star formation history or the metallicity~\cite{nat.460.990.09,prd.79.062002.09,prd.80.122002.09,raap.11.369.11,aap.574.A58.15,prl.109.171102.12}.

In order to prepare and test our ability at detecting the SGWB and interpreting valuable information from the data, we are conducting a series of mock data and science challenges (MDSC), with increasing degrees of complexity.
In this first paper, we focus on a SGWB created by all the unresolvable compact binary coalescences (CBC) such as binary neutron stars (BNS), neutron star-black holes (NSBH) or binary black holes (BBH), up to a redshift $z$ = 10, which may dominate within our search frequency range.
Such a background may have a realistic chance of being detected after a few years of operation of the advanced detectors~\cite{prd.85.104024.12}.
The observation of the SGWB will complement individual detections of a few to a few tens of CBC events per year~\cite{cqg.27.173001.10} at close distances up to a few hundred Mpc.

For this study we produced multiple year-long data sets in the form of time series for the three advanced LIGO/Virgo detectors, containing both instrumental noise and the GW signal from a large number ($10^4 - 10^7$ per year) of compact binary sources out to redshift $z$~=~10.
This is done using a data generation package that was initially developed for the Einstein Telescope MDSC~\cite{prd.86.122001.12, prd.89.084046.14}.
The data sets are then analysed using a cross-correlation (CC) analysis pipeline in order to measure the total GW energy density, $\Omega_{\mathrm{gw}}(f)$, of all the GW signals that make up the stochastic background~\cite{prl.89.231101.14}.
We then use the results from these analyses to perform parameter estimation to try to determine some of the parameters of the injected populations, such as the average mass of all the sources, or the rate at which these binaries coalesce.

This paper is organised into the following sections.
In Section~\ref{sec:mockdata}, we introduce the mock data sets and the methods by which we generate them.
In Section~\ref{sec:analysis}, we briefly describe the analysis methods used to detect the stochastic signal.
In Section~\ref{sec:PE}, we discuss how we use the results from the analysis to estimate various astrophysical parameters.
In Section~\ref{sec:results}, we present the results from various analysis runs of the different mock data sets.
Finally in Section~\ref{sec:conclusion}, we present our conclusions.


\section{Mock Data}
\label{sec:mockdata}

In this section, we introduce the mock data sets that we will be analysing as part of this investigation, as well as the data generation program we use to produce them.
Initially we explain the various steps that are used to produce the mock data before detailing each of the data sets that will be considered as part of this MDSC.
Finally we show how one can consider an astrophysical SGWB as the superposition of many unresolvable single sources.
The main data sets are split into two subsets, the first being produced with Gaussian noise, and the second produced using ``glitchy''  data taken from the initial LIGO S5 and initial Virgo VSR1 science runs re-coloured to have the sensitivity of the advanced detectors~\cite{prd.82.102001.10}.
The GW signals injected into each set are the same for both the Gaussian and re-coloured noise, allowing us to make a direct comparison of how the analysis pipeline will behave in an ideal and in a more realistic case.


\subsection{Mock Data Generation}

\begin{figure}
\hskip -0.3cm
\includegraphics[width=0.45\textwidth]{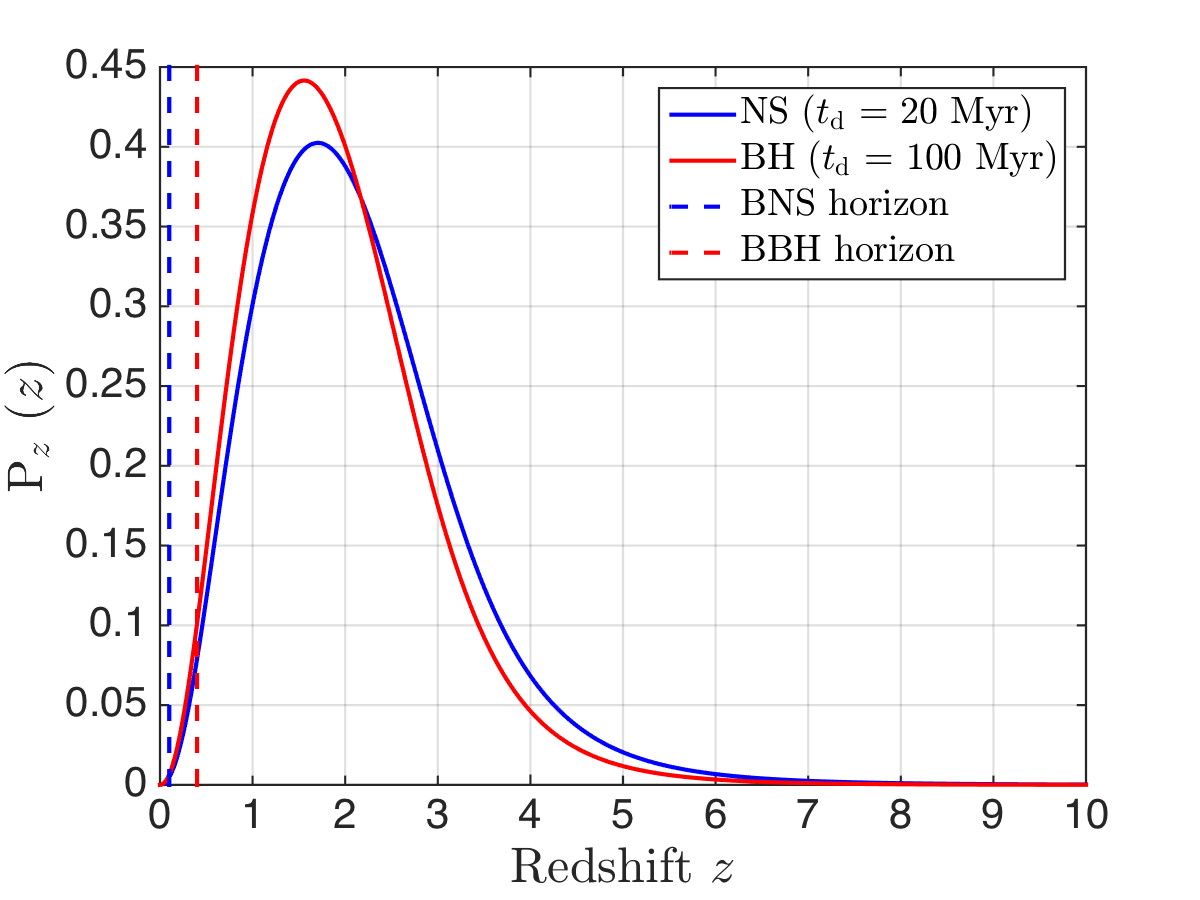}
\caption{Source redshift probability distribution for BNS (blue) and BBH (red) where different delay times between the formation and merger of the binary system are considered.
We also plot the horizon distance to BNS (dashed blue) and BBH (dashed red) for the advanced detectors with their design sensitivity which are defined as 445~Mpc ($z \sim 0.1$) and 2187~Mpc ($z \sim 0.4$) respectively~\cite{cqg.27.173001.10}.}
\label{fig:Pz}
\end{figure}

The mock data generation package used here was originally developed for the Einstein Telescope MDSC~\cite{prd.86.122001.12} where one would expect to be able to make detections of individual sources out to $z \approx 3.8$ for BNS  and even further for higher mass systems such as NSBH or BBH.
Being able to realistically represent the population of sources at high redshift is essential when considering an astrophysical SGWB signal for the advanced detectors.
This is because we expect very few CBC events to be directly detectable but the large number of unresolvable sources, when considering the whole universe, will all contribute to the SGWB.
This is clearly seen in Fig.~\ref{fig:Pz} where we show the redhsift probability distribution, which is explained in the next section, of BNS (blue) and BBH (red) which take into account different delay times between the formation and merger of the inspiralling systems.
We also plot the maximum horizon distance of the advanced detectors to directly detect individual BNS (blue dashed) and BBH (red dashed) signals~\cite{prd.79.122001.09}.
We now describe how we generate and add the large number of GW signals to the detector data streams using both Gaussian and re-coloured noise.

\begin{figure*}
\hskip -0.3cm
\includegraphics[width=0.45\textwidth]{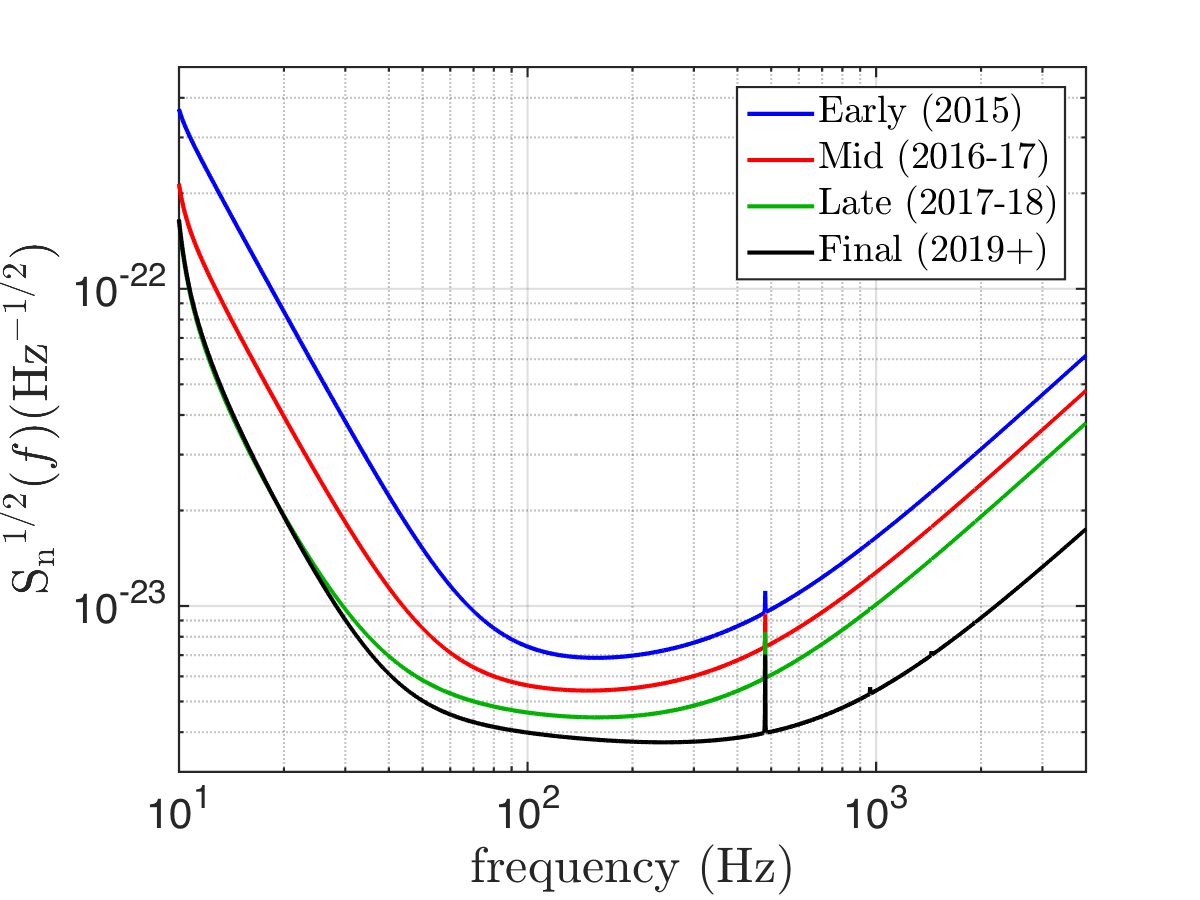}
\hskip -0.5cm
\includegraphics[width=0.45\textwidth]{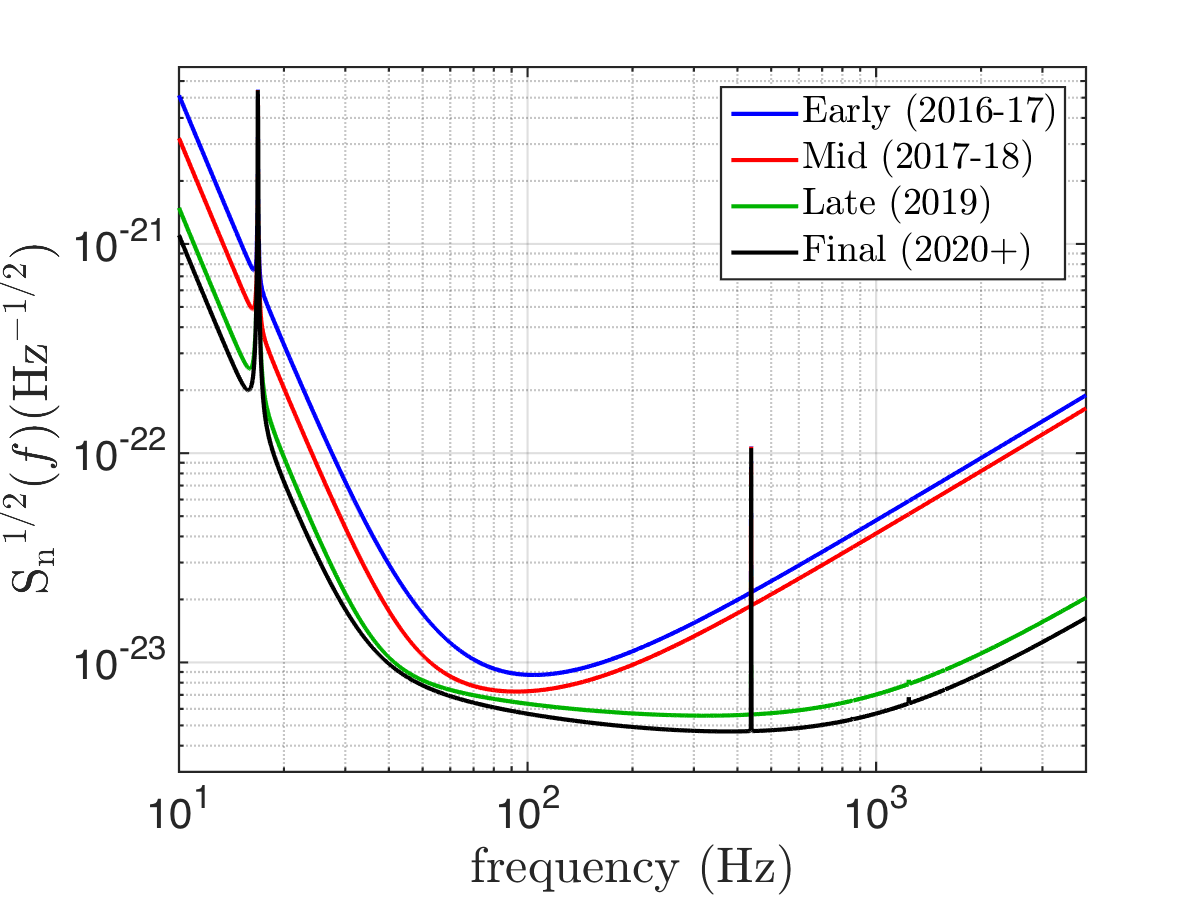}
\caption{\emph{Left}--- Evolution of the Advanced LIGO sensitivities over the early (blue), middle (red), late (green) and design (black) phases~\cite{ObsScenario}.
\emph{Right}--- Evolution of the Advanced Virgo sensitivities over the early (blue), middle (red), late (green) and design (black) phases~\cite{ObsScenario}.}
\label{fig:noise}
\end{figure*}

\subsubsection{Simulation of GW CBC signals}
\label{sec:cbcsim}

The Monte Carlo procedure we use to generate a population of compact binaries up to redshift of $z=10$ is described in detail in~\cite{prd.86.122001.12,apj.799.69.15}.
Here we summarize the main steps of the simulations.

The coalescence of a compact binary occurs after two massive stars in a binary system have collapsed to form a neutron star or a black hole\footnote
{We neglect the possible production of compact binaries through interactions in dense star systems.}
and have inspiralled through the emission of gravitational waves.

The coalescence rate in Mpc$^{-3}$ yr$^{-1}$ is given by~\cite{apj.648.1110.06,apj.664.1000.07,apjl.550.L183.01,aap.415.407.04}

\begin{equation}
\dot{\rho}_c(z,t_d)= \lambda \int \frac{\dot{\rho}_*(z_f)}{1+z_f} P(t_d) \dif t_d .
\end{equation}

\noindent In this expression, $\dot\rho_*$ is the star formation rate (SFR), measured in M$_\odot$ Mpc$^{-3}$ yr$^{-1}$ and the factor $1/(1+z_f)$ accounts for time dilation due to the cosmic expansion.
The redshift $z$ describes when our compact binary merges, $z_f$ is the redshift at which its progenitor massive binary formed, and $P(t_d)$ is the probability distribution of the delay between $z$ and $z_f$, which is the sum  of the time from initial binary formation to evolution into a compact binary, plus the merger time by emission of gravitational waves.
The parameter $\lambda$ (in M$_\odot^{-1}$) is the mass fraction that goes into the formation of the progenitors of compact binaries.
The local coalescence rate at $z=0$, $\dot{\rho}_c(0,t_d)$, is one of the parameters of our simulations and is denoted by $\rho_0$.

The merger rate in the redshift interval $[z, z+\dif z]$ is obtained by multiplying $\dot{\rho}_c(z,t_d)$ with the element of comoving volume

\begin{equation}
\frac{\dif R}{\dif z}(z,t_d)=\dot{\rho}_c(z,t_d) \frac{\dif V}{\dif z}(z) ,
\label{eq-dRdz}
\end{equation}

\noindent where

\begin{equation}
\frac{\dif V}{\dif z}(z)=4 \pi \frac{c}{H_0} \frac{r(z)^2}{E(\Omega,z)},
\label{eq-dVdz}
\end{equation}

\noindent where $c$ is the speed of light in a vacuum, $H_0$ is the Hubble constant and

\begin{equation}
r(z)= \frac{c}{H_0}\int_0^z \frac{\dif z'}{E(\Omega, z')},
\label{eq-r_z}
\end{equation}

\noindent where

\begin{equation}
E(\Omega,z)=\sqrt{\Omega_{\Lambda}+\Omega_{m}(1+z)^3}\;.
\label{eq-E-z}
\end{equation}

Here, we select the SFR given in~\cite{apj.651.142.06} and use a standard $\Lambda$CDM cosmology with $\Omega_m=0.3$, $\Omega_{\Lambda}=0.7$ and Hubble parameter $H_0=70$ km s$^{-1}$ Mpc$^{-1}$.
Following~\cite{prd.79.062002.09,prd.85.104024.12} we assume a distribution of the delay of the form $P(t_d) \propto 1/t_d$ with a minimal delay of 20 Myr for BNS and 100 Myr for BBH, as suggested by population synthesis~\cite{apjs.174.223.08,apj.759.52.12,mnras.413.461.11}.

We then proceed as follows for each source:

\begin{itemize}
\item The arrival time $t_c$ of each GW signal is selected from a Poisson distribution.
Here, the difference in arrival time, $\tau = t_{c}^{k} - t_{c}^{k-1}$, where $k$ is the current event, is drawn from the exponential distribution $P(\tau) = \exp(-\tau/\overline{\tau})$ where $\overline{\tau}$ is the average time between successive events.
The average waiting time between signals is calculated by taking the inverse of the coalescence rate, Eq.~(\ref{eq-dRdz}), integrated over all redshifts

\begin{equation}
\overline{\tau} = \left[ \int_0^{10} \frac{\dif R}{\dif z} (z,t_d) \, \dif z \,\right] ^{-1} .
\end{equation}

\item The redshift at the point of coalescence, $z$, is selected from a probability distribution $p(z,t_d)$ constructed by normalizing the coalescence rate in the interval $[0, 10]$ (see Fig.~\ref{fig:Pz})

\begin{equation}
p(z,t_d) = \overline{\tau} \frac{\dif R}{\dif z}(z,t_d) .
\end{equation}

\item The SGWB analysis is not sensitive to the width of the distribution of the masses, only to the average chirp mass, $\mathcal{M}$, of the system.
This is a combination of the two component masses, $m_1$ and $m_2$, given by

\begin{equation}
\mathcal{M} = \frac{(m_1 m_2)^{3/5}}{(M)^{1/5}},
\end{equation}

where $M = m_1 + m_2$ is the total mass of the system.
Because of this, we choose a single value for the component masses for each of the systems being considered: 1.4M$_\odot$ for neutron stars and 10M$_\odot$ for black holes.

\item The sky position, $\hat{\Omega}$, is selected from an isotropic distribution across the whole sky.
The cosine of the inclination angle of the orbital plane to our line of sight, $\iota$, the GW polarisation angle, $\psi$, and the phase angle at the time of coalescence, $\phi_0$, are all chosen from uniform distributions.

\item We next calculate the theoretical signal-to-noise ratio (SNR), $\rho$, of the inspiral signal to determine if it is individually detectable by the standard LIGO-Virgo CBC search pipeline~\cite{prd.79.122001.09,prd.80.047101.09,prd.82.102001.10,prd.85.082002.12}.
The SNR produced by the inspiral phase of the waveform for CBCs is given by

\begin{equation}
\rho^2 = 4 \int_{f_\mathrm{min}}^{f_\mathrm{lsco}} \frac{ \left| F_+(\hat{\Omega},\psi)\tilde{h}_+(f) + F_\times(\hat{\Omega},\psi) \tilde{h}_\times(f) \right|^2}{S_n(f)}\, \dif f , 
\end{equation}

where $F_+$ and $F_\times$ are the antenna response functions to the two GW polarisations originating from sky position $\hat{\Omega}$ and with polarisation angle $\psi$, and $\tilde{h}_+$ and $\tilde{h}_\times$ are the signal amplitudes in the frequency domain for the two polarisations.
In the Newtonian regime, before the last stable circular orbit,  $\tilde{h}_+$ and $\tilde{h}_{\times}$ are given by

\begin{eqnarray}
\tilde{h}_+(f) & = & h_z \, \frac{(1+\cos^2\iota)}{2} \, f^{-7/6}, \label{eq:fourierhp} \\
\tilde{h}_\times(f) & = & h_z \, \cos\iota \, f^{-7/6}, \label{eq:fourierhc}
\end{eqnarray}

\noindent where

\begin{equation}
h_z = \sqrt\frac{5}{24}\,\frac{(G \mathcal{M}(1+z))^{5/6}}{\pi^{2/3} c^{3/2} d_{\rm L}(z)} .
\label{eq:hz}
\end{equation}

In the above equations $G$ is the gravitational constant, $d_\mathrm{L}$ is the luminosity distance to the source at redshift $z$, $f_\mathrm{min}$ is the starting frequency, which we select to be 10\,Hz, $S_n$ is the detector's noise power spectral density (PSD) (see Fig.~\ref{fig:noise}), and $f_\mathrm{lsco}$ is the frequency of the last stable circular orbit, $f_\mathrm{lsco}~\simeq~\dfrac{c^3}{6^{3/2} \pi G M}$.
For BNS signals it is enough to consider the waveform up until this point as the SNR contribution of the inspiral phase is dominant.
However, for BBH signals we must also consider the contribution from the merger and ringdown of the waveform.
The modifications to the calculations of $\tilde{h}_+$ and $\tilde{h}_{\times}$ are given in~\cite{prd.77.104017.08}.
The total SNR for the GW detector network is then given by

\begin{equation}
\rho^2 = \sum_A \rho_A^2 ,
\end{equation}

where $A$ is the sum over all detectors in the network.
Any signals that pass a network threshold SNR $\rho_\mathrm{T}$ are then ignored by the SGWB search, where we set the network SNR threshold to 12.

\item Finally, for any surviving sub-threshold events, we produce the waveforms that are then added to the detector data steams.
Here, we have chosen to use the TaylorT4 time-domain waveform up to 3.5 post-Newtonian order in phase, and the most dominant post-Newtonian lowest order for amplitude, for the BNS signals.
For the case of BBH signals, we choose the EOBNRv2 waveform produced from numerical relativity, which includes the merger and ringdown of the two coalescing black holes.
This is up to 4th post-Newtonian order for phase and the lowest order for amplitude~\cite{prd.80.084043.09}.
\end{itemize}

\noindent Once the time series data has been produced containing all the sub-threshold injections, we add either Gaussian noise or re-coloured noise data to produce the final mock data sets.

\subsubsection{Simulation of Gaussian noise}

Because each of the detectors being considered in this MDSC are well-separated in space, we assume that there will be no correlated noise between any of them; so the noise is simulated independently for each of the detectors~\cite{prd.87.123009.13,prd.90.023013.14}.
We do this by generating a mean zero, unit variance Gaussian time series which is then Fourier transformed into the frequency domain.
This is then coloured using the PSD of the detector sensitivity of either aLIGO or AdVirgo (see Fig.~\ref{fig:noise}) and is then finally Fourier transformed back into the time domain.
To prevent any potential discontinuities of the data between adjacent segments of data, we taper the noise curve away to zero at frequencies below 10Hz and above 512Hz.
This is then added to the time series containing the injected GW signals.

\subsubsection{Re-coloured noise}

To more accurately imitate the noise likely to be present in the advanced detectors, we also re-colour initial S5 LIGO and Virgo VSR1 data to have the sensitivity of aLIGO and AdVirgo.
This has the benefit of preserving non-stationary noise artifacts while exploring the sensitivity of the pipelines to the signals.
The noise spectra are estimated from the year of data and then averaged.
The data is then re-coloured with a transfer function corresponding to the advanced detector power spectra divided by this averaged spectrum.
The same GW signals are then added to these re-coloured time-series.


\subsection{Mock Data Sets}
\label{sec:dataSets}

To ensure that we have a detectable signal and to reduce computational costs, we have selected astrophysical models corresponding to the most optimistic rates given in~\cite{cqg.27.173001.10} instead of using longer observational times to obtain the same SNR.

In total, we generate 5 data sets, of duration one year, all of which are produced with both Gaussian noise and re-coloured noise.
For these data sets, we use the nominal design sensitivities of Advanced LIGO and Virgo, but in order to account for the improvement of the sensitivity, we also produce a sixth, observing scenario data set consisting of 3, 6, and 9 months and 1 and 3 years corresponding to the early, middle, late and design phases of the advanced detectors (see Fig.~\ref{fig:noise}).
This observing scenario, consisting of 5.5 years of data, should cover the full advanced detector observing period from mid 2015 until the end of 2022.
Details of all the data sets are found in Table~\ref{tab:DataSets} and details on the rates and expected number of events are found in Table~\ref{tab:Rates}.

One part of the investigation is to see how having statistically different sources effects our analysis.
When the rate and the duration of the GW events are large, the sources overlap each other creating a GW signal continuous in time (there is always a source present) which is Gaussian in nature due to the central limit theorem.
However, for smaller rates or shorter waveforms, the time interval between successive events increases resulting in a non-continuous and non-Gaussian signal~\cite{nar.50.461.06,prd.79.062002.09}.
Analytical calculations made in~\cite{prd.89.084063.14} and for the Einstein Telescope MDSCs~\cite{prd.86.122001.12,prd.89.084046.14} suggest that when making a measurement of an astrophysical gravitational-wave background one just needs to consider the total number of coalescing events, with their relevant signal amplitudes, that occur within the observational period. 
The nature of the signals themselves will have no effect on the estimation of the SGWB.
This is explained in greater detail later in Section~\ref{sec:CBC}.
However these results have not been independently verified with the use of simulated data.
We now describe each of these data sets being considered in this investigation.

\subsubsection{Main data sets}

Data set 0 is our control test for both the Gaussian and re-coloured data sets.
The data streams for each of the detectors contain no coincident signals so there will be no correlated signals between any of the detectors.
Thus, the results from the analysis of data set 0 should give us an accurate measurement of the expected error bars for each of the following data sets.

\begin{figure}
\hskip -0.3cm
\includegraphics[width=0.45\textwidth]{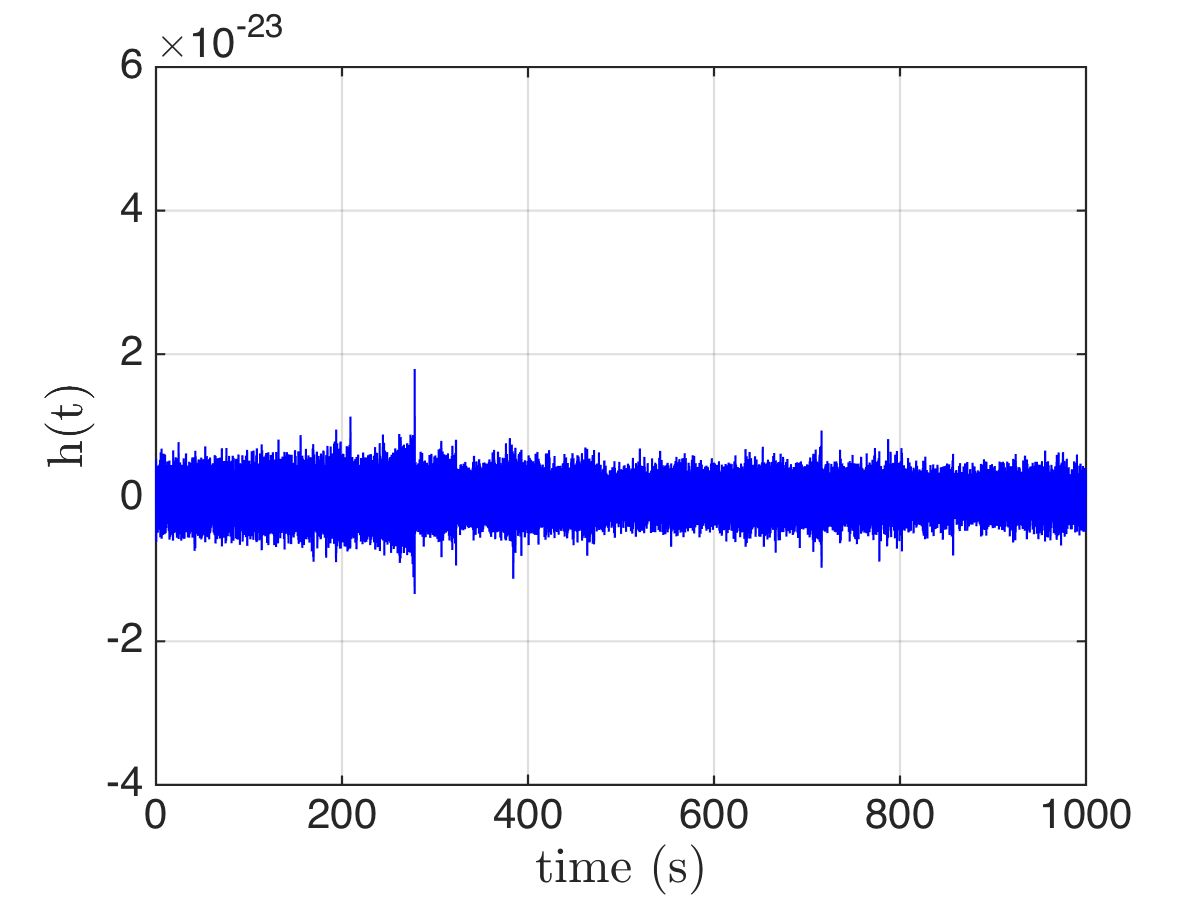} \\
\hskip -0.3cm
\includegraphics[width=0.45\textwidth]{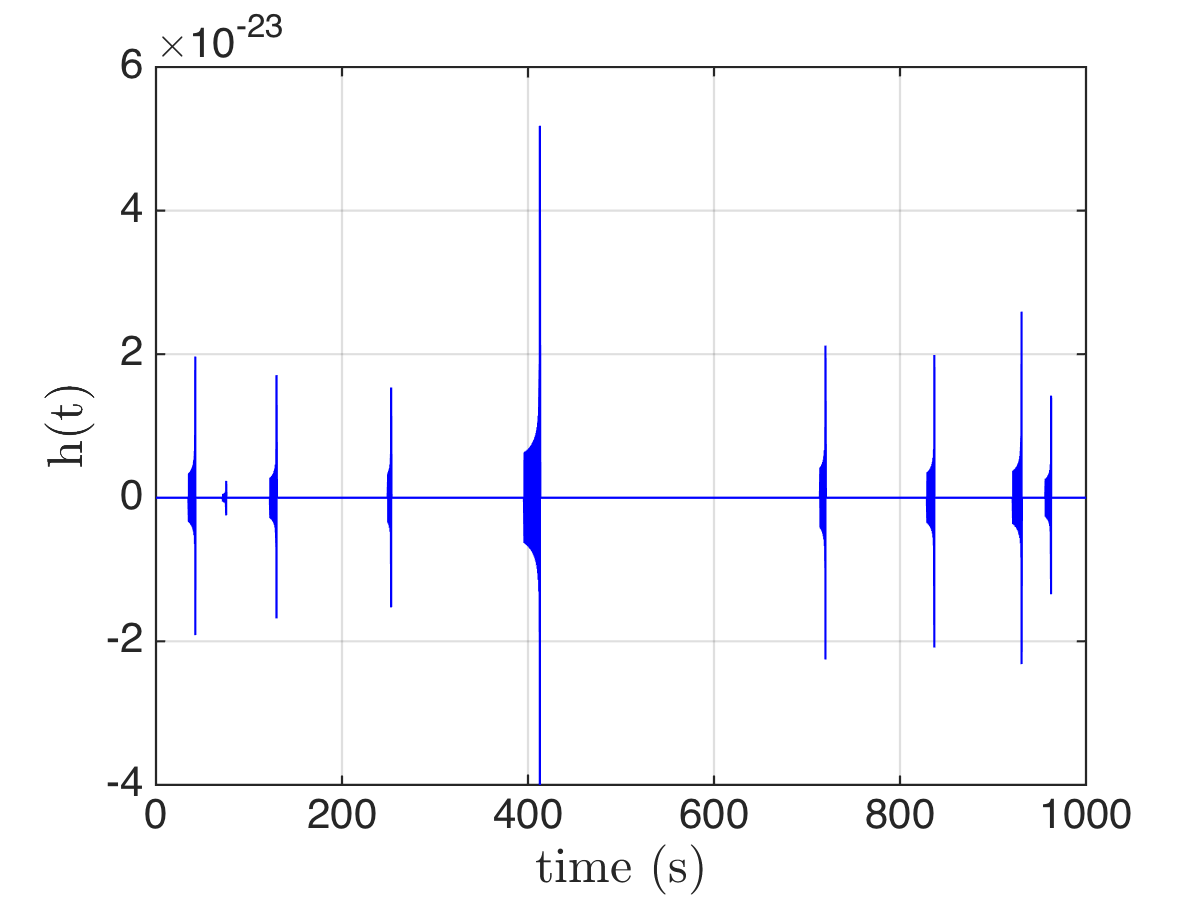} \\
\hskip -0.3cm
\includegraphics[width=0.45\textwidth]{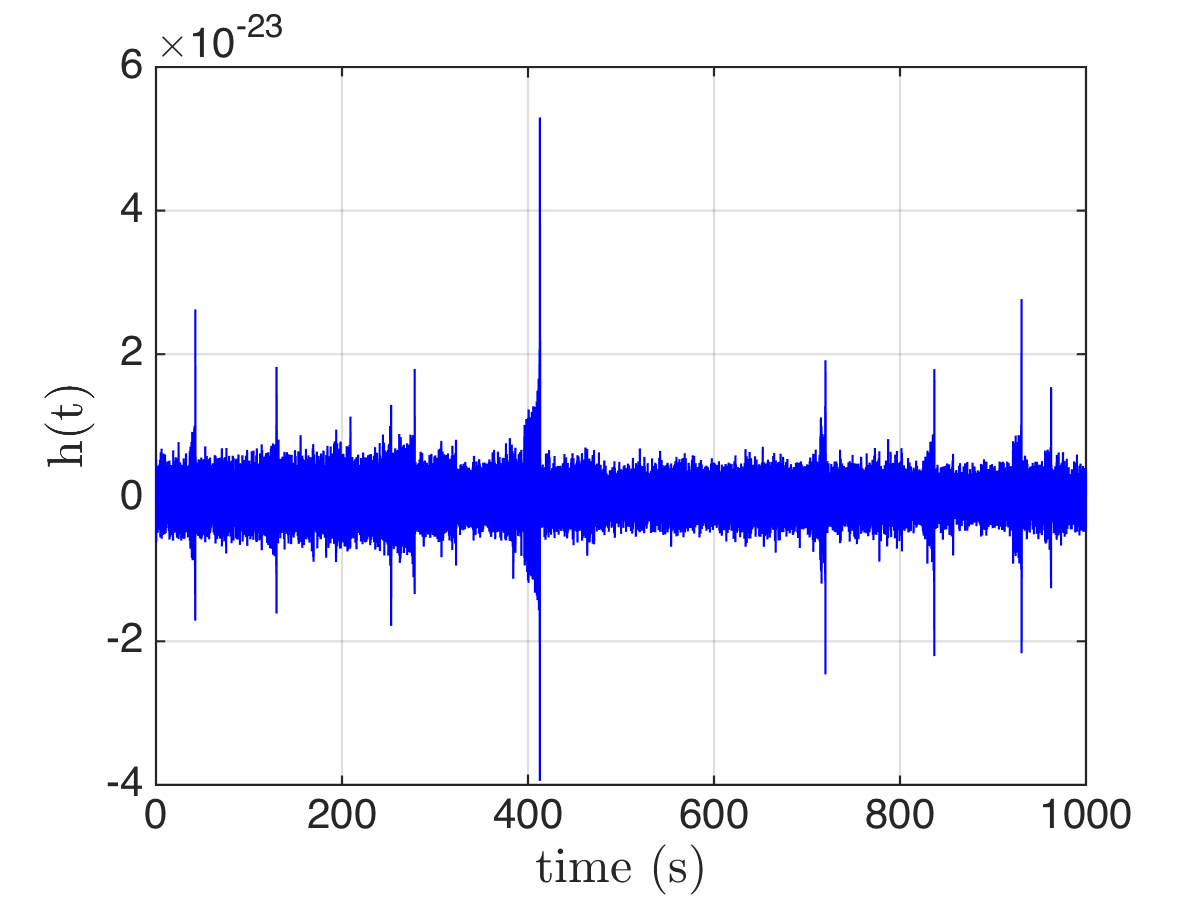} \\
\caption{\emph{Top}--- A 1000s segment of the time series for BNS signals using the higher LIGO rate of 10 Mpc$^{-3}$ Myr$^{-1}$.
This signal appears continuous.
\emph{Middle}--- A 1000s segment of the time series for BBH signals of mass $10+10$ M$_{\odot}$, using the higher LIGO rate of 0.3 Mpc$^{-3}$ Myr$^{-1}$.
This signal appears non-continuous (popcorn).
\emph{Bottom}--- A 1000s segment of the time series for mixed BNS and BBH signals with higher LIGO rates and BH mass of 10M$_{\odot}$.
This signal appears as a popcorn background from BBH on top of a continuous background from BNS.}
\label{fig:timeSeries}
\end{figure}

In data set 1, we have generated a large number of BNS signals with a merger rate of 10 coalescences per Mpc$^3$ per Myr that are injected into both the Gaussian and the re-coloured noise.
Any individual events that surpasses a network SNR threshold value, as described in Section~\ref{sec:cbcsim}, are removed as it is possible for these signals to bias the results of the analysis.
But given the expected number and length of the waveform when compared to the overall length of the observing time this effect is negligible, as demonstrated later in this paper.
The top plot of Fig.~\ref{fig:timeSeries} shows a 1000s segment of the time series produced from BNS contributing to data set 1.
It is clearly seen that the GW signal is continuous at all times, so with this data set we investigate the affect that a continuous SGWB signal will have on our analysis.

Data set 2 contains the exact same sources with both Gaussian and re-coloured noise that are included in set 1 except that here we have not removed the loudest individually detectable events from the time series.
This is to test by how much our results can be affected if we include the loud events in the analysis by comparing the results against that of the first data set.

In data set 3, we generate a number of BBH signals using a merger rate of 0.3 coalescences per Mpc$^3$ per Myr that are injected into both Gaussian and re-coloured data.
With this data set we investigate the possible effects that a non-continuous (popcorn) SGWB will have on our analysis.
The middle plot of Fig.~\ref{fig:timeSeries} shows a 1000s segment of the time series produced from BBH contributing to data set 3.
We see that, due to the shorter waveform lengths and lower coalescence rate, the GW signals are non-continuous or more ``popcorn" like.

In data set 4 we have generated a number of both BNS and BBH signals, using the same rates stated above, which are injected into both Gaussian and re-coloured data.
This set is to test the behaviour of the analysis and parameter estimate when analysing data from a SGWB having a contribution from more than one source.
The bottom plot of Fig.~\ref{fig:timeSeries} shows a 1000s segment of the time series produced from BNS and BBH signals contributing to data set 4.
It is clearly seen that the GW signal is continuous at all times whilst still having louder popcorn like bursts from the BBH signals.

\subsubsection{Observing scenario}

The observing scenario data set is designed to realistically simulate the data that we would expect to obtain from the network of advanced detectors over the initial several years of operations.
For this set, we generate a large number of BNS signals using a merger rate of 2 coalescence per Mpc$^3$ per Myr that are injected into Gaussian data.
This is a lower rate than is used in set 1 as we are using a longer observational period.
This Gaussian data differs from the previous data sets as we change the PSDs for each of the detectors at different stages to represent the improvements in sensitivities that are expected to be obtained in each observing run.
Examples of this are shown in the left-hand plot for aLIGO, and right-hand plot for AdVirgo of Fig.~\ref{fig:noise}.
In reality we should take the final sensitivity of one phase as the initial sensitivity of the next, which would then gradually decrease to the next final sensitivity.
However, here we consider the ideal case of taking the final sensitivity for the full duration of each phase.
These observing runs are set out as follows~\cite{ObsScenario}:

\begin{enumerate}
\item O1, 2015: This will consist of a 3-month observational run with both LIGO detectors (HL) with Early aLIGO sensitivity.
\item O2, 2016-17: This will consist of a 6-month observational run with all three detectors (HLV), where the HL detectors will have Mid aLIGO sensitivity and V will have  Early AdVirgo sensitivity.
\item O3, 2017-18: This will consist of a 9-month observational run with all three detectors (HLV), where the HL detectors will have Late aLIGO sensitivity and V will have  Mid AdVirgo sensitivity.
\item O4, 2019: This will consist of a year long observational run with all three detectors (HLV), where the HL detectors will have the final design aLIGO sensitivity and V will have Late AdVirgo sensitivity.
\item O4 (\textit{continued}), 2020-22: This will consist of a 3 year long observational run with all three detectors (HLV), where the HL detectors will have the final design aLIGO sensitivity and V will have the final design AdVirgo sensitivity.
\end{enumerate}

\begin{table*}
\caption{\label{tab:DataSets} Table describing the data sets that are produced as part of the MDSC.
The first column is the reference number of the data set.
The second column indicates if the data set is produced with just Gaussian noise or with re-coloured noise as well.
The third column shows what sources are injected into the data set.
The forth column gives the rate of events used, see Table~\ref{tab:Rates}.
The fifth column gives the length of the data set.}
\begin{center}
\begin{tabular}{ccccc}
\hline \hline
Data Set & Noise & Sources & Rate & $\mathrm{T_{obs}}$ \\ \hline
0 & Gaussian \& Re-coloured & -- & -- & 1 year \\
1 & Gaussian \& Re-coloured & BNS (sub-threshold) & 10 $\mathrm{Mpc^{-3} Myr^{-1}}$ & 1 year \\
2 & Gaussian \& Re-coloured & BNS (all) & 10 $\mathrm{Mpc^{-3} Myr^{-1}}$ & 1 year \\
3 & Gaussian \& Re-coloured & BBH (sub-threshold) & 0.3 $\mathrm{Mpc^{-3} Myr^{-1}}$ & 1 year \\
4 & Gaussian \& Re-coloured & BNS + BBH (sub-threshold) & 10+0.3 $\mathrm{Mpc^{-3} Myr^{-1}}$ & 1 year \\
Observing Scenario & Gaussian & BNS (sub-threshold) & 2 $\mathrm{Mpc^{-3} Myr^{-1}}$ & 5.5 years \\
\hline \hline
\end{tabular}
\end{center}
\end{table*}

\begin{table}
\caption{\label{tab:Rates} A list of compact binary coalescence rate densities  as given in~\cite{cqg.27.173001.10}.
The first column labels whether a merger rate is optimistic ($R_\text{high}$), realistic ($R_\text{realistic}$) or pessimistic ($R_\text{low}$).
The second column gives the rates of coalescing events. 
The third column gives the average time between successive events.
The final column gives the total number of events out to $z=10$ that are expected to occur per year.}
\begin{center}
\begin{tabular}{ l c c c }
\hline \hline
Expected Rate & $\dot{\rho}_0$ (Mpc$^{-3}$ Myr$^{-1})$ & $\overline{\tau}$ (s) & N$_\mathrm{events}$ (yr$^{-1}$) \\ \hline
\multicolumn{4}{c}{BNS} \\ \hline
$R_\text{high}$ & 10 & 1.35 & $2.3\times10^7$ \\
$R_\text{realistic}$ & 1 & 13.5 & $2.3\times10^6$ \\
$R_\text{medium-low}$ & 0.1 & 135 & $2.3\times10^5$ \\
$R_\text{low}$ & 0.01 & 1350 & $2.3\times10^4$ \\ \hline
\multicolumn{4}{c}{BBH} \\ \hline
$R_\text{high}$ & 0.3 & 64.7 & $4.9\times10^5$ \\
$R_\text{realistic}$ & 0.005 & 3880 & $8133$ \\
$R_\text{medium-low}$ & 0.001 & 19400 & 1627 \\
$R_\text{low}$ & 0.0001 & 194000 & 163 \\  \hline \hline
\end{tabular}
\end{center}
\end{table}

\subsection{Astrophysical stochastic background from CBC}
\label{sec:CBC}

The spectrum of the SGWB is usually characterized by the dimensionless parameter

\begin{equation}
\Omega_{\mathrm{gw}}(f)=\frac{1}{\rho_c}\frac{\dif\rho_{\mathrm{gw}}}{\dif \ln f},
\label{eq:omega_def}
\end{equation}

\noindent where $\rho_{\mathrm{gw}}$ is the gravitational-wave energy density, $f$ the frequency in the observer's frame and, $\rho_c=\frac{3c^2H_0^2}{8 \pi G}$, is the critical energy density needed to close the Universe today.
The GW spectrum from the population of extra-galactic compact binaries is given by the expression

\begin{equation}
\Omega_{\rm{gw}}(f)=\frac{1}{\rho_c \, c} f F(f),
\label{eq:omega_flux}
\end{equation}

\noindent where $F(f)$ is the total flux.
The total flux is the sum of the individual contributions

\begin{equation}
F(f)= T_\mathrm{obs}^{-1}  \frac{\pi c^3}{2G} f^2 \sum_{k=1}^{N} (\tilde{h}^2_{+,k} + \tilde{h}^2_{\times,k}) ,
\label{eq:flux}
\end{equation}

\noindent where $N$ is the total number of coalescences in the data and $k$ is the index of the individual coalescence.
The normalization factor $T_\mathrm{obs}^{-1}$ assures that the flux has the correct dimension, $T_\mathrm{obs}=1$ yr being the length of the data sample.

In the Newtonian regime, before the last stable circular orbit, the Fourier transforms $\tilde{h}_+$ and $\tilde{h}_{\times}$ are given by Eqs.~(\ref{eq:fourierhp}~-~\ref{eq:hz}).
This gives for the energy density parameter~\cite{prd.89.084063.14}

\begin{widetext}
\begin{equation}
\Omega_{\rm{gw}}(f) = \frac{5 \pi^{2/3} G^{5/3}c^{5/3}}{18 c^3 H_0^2}  f^{2/3} \sum_{k=1}^{N} \frac{ \left(\mathcal{M}_k(1+z_k) \right)^{5/3}}{d_{\rm L}(z_k)^2} \left(\frac{(1+\cos^2 \iota_k)^2}{4}+\cos^2\iota_k \right) .
\label{eq:omega_discrete}
\end{equation}
\end{widetext}

\begin{figure*}
\hskip -0.3cm
\includegraphics[width=0.45\textwidth]{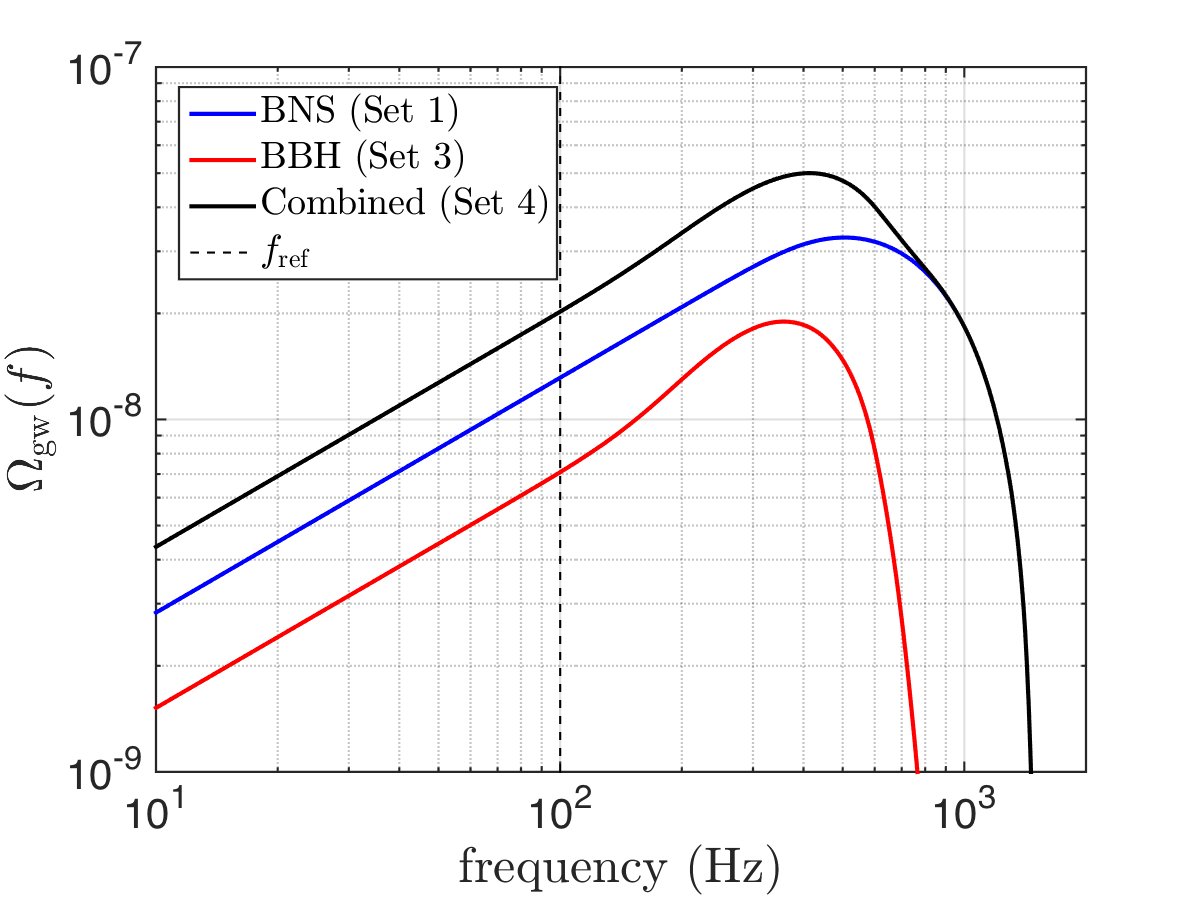}
\hskip -0.5cm
\includegraphics[width=0.45\textwidth]{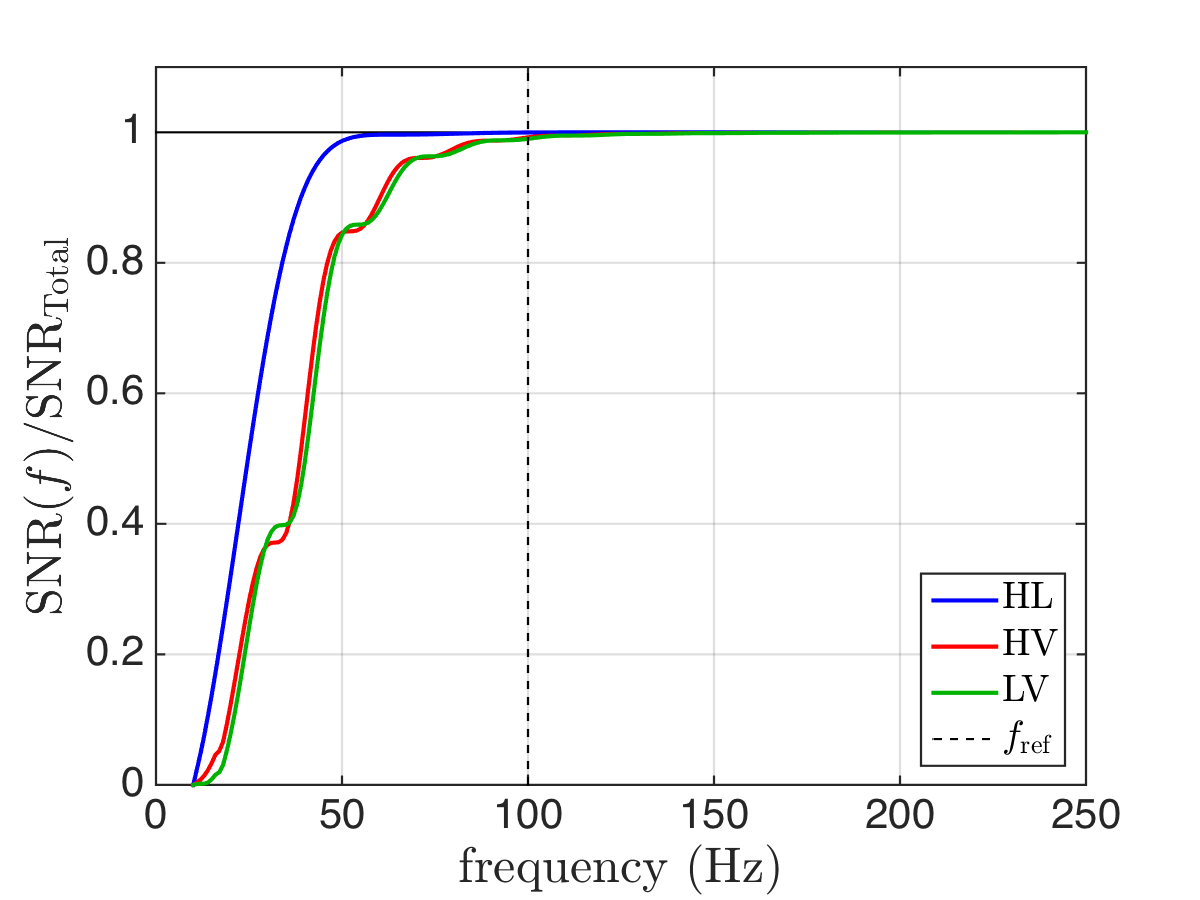}
\caption{\emph{Left}--- Energy density $\Omega_{\rm gw}(f)$ for the population of BNS (blue), BBH (red), and the combination of the two populations (black) as used in data sets 1, 3 and 4 respectively.
The plots are calculated using the higher LIGO rates of 10 Mpc$^{-3}$ Myr$^{-1}$ (BNS) and 0.3 Mpc$^{-3}$ Myr$^{-1}$ (BBH).
We also plot the reference frequency for which we report all the results in this investigation, shown by the black dashed line.
\emph{Right}---  Fraction of the total theoretical SNR for the three detector pairs, HL (blu), HV (red), and LV (green).
This is calculated using the frequency band $[10-250]$Hz, as given by Eq.~(\ref{eq:ccSNR}), and assuming $f^{2/3}$, which is used for the analysis.
We also plot the reference frequency for which we report all the results in this investigation, shown by the black dashed line.}
\label{fig:background}
\end{figure*}

\noindent This equation is valid for BNS signals where we have considered only the inspiral phase, but for BBH signals, there is an extra contribution coming from the merger and ringdown phases.
Fig.~\ref {fig:background}  shows  $\Omega_{\mathrm{gw}}(f)$ for the population of BNS (blue), and BBH (red) in the mock data sets described in the previous section.
The plot in black corresponds to the sum of the signal from BNS and BBH.
For BNS and BBH signals, $\Omega_\mathrm{gw}(f)$ increases as $f^{2/3}$ from the inspiral phase (then as $f^{5/3}$ from the merger phase for BBH) before it reaches a maximum and decreases dramatically.
The peaks occur at frequencies corresponding roughly to the $f_{\mathrm{lsco}}$ and the end of the ring-down phase at $z \sim 1.5$ where the coalescence rate is maximal.
The amplitude of the background scales with both the rate of coalescing events and the average chirp mass of all the signals.
It is larger for the BNS background (data set 1) than for the BBH contribution (data set 3) because, even though the chirp mass is smaller, the rate we have considered is larger.

\section{SGWB Search}
\label{sec:analysis}

In this section, we briefly describe the cross-correlation (CC) method by which we analyse the data (please see references such as~\cite{prd.59.102001.99,nat.460.990.09,prl.89.231101.14} for a complete treatment).
The optimal strategy to search for a Gaussian (or continuous) SGWB is to cross-correlate measurements of multiple detectors, $\tilde{s_1}(f)$ and $\tilde{s_2}(f)$.
When the background is assumed to be isotropic, unpolarized and stationary, the cross correlation product is given by~\cite{prd.59.102001.99}

\begin{equation}
Y \simeq \int_{-\infty}^\infty \tilde{s_1}^*(|f|) \tilde{s_2}(|f|) \tilde{Q}(f) \dif f ,
\end{equation}

\noindent and the expectation value of $Y$ is

\begin{equation}
<Y> = \frac{3H_0^2}{20 \pi^2} T_{\mathrm{obs}} \int_{-\infty}^\infty \frac{1}{|f|^3} \Omega_{\rm gw}(|f|) \gamma(|f|) \tilde{Q}(f) \dif f ,
\label{eq:ccstat}
\end{equation}

\noindent where

\begin{equation}
\tilde{Q}(f)\propto \frac{\gamma (f) \Omega_{\mathrm{gw}}(|f|)}{|f|^3P_1(|f|)P_2(|f|)} ,
\end{equation}

\noindent is the optimal filter that maximizes the SNR, $\Omega_{\rm gw}(f)$ is the energy density in GW as defined in Eq.~(\ref{eq:omega_def}), $\tilde{s_1}$ and $\tilde{s_2}$ are the detector output streams from both detectors in the frequency domain, $P_1(f)$ and $P_2(f)$ are the detector power spectral densities of the two detectors and $\gamma(f)$ is the normalized overlap reduction function~\cite{prd.46.5250.92}, characterizing the loss of sensitivity due to the separation and the relative orientation of the detectors.

The expected variance, which is dominated by the noise, is given by

\begin{equation}
\sigma^2_Y \approx  \frac{T_{\mathrm{obs}}}{4} \int_{-\infty}^\infty P_1(|f|)P_2(|f|)|\tilde{Q}(f)|^2 \dif f ,
\label{eq-ccvar}
\end{equation}

\noindent and the expected (power) SNR of the CC statistic in the frequency range $f_{\min}-f_{\max}$, for an integration time $T_{\mathrm{obs}}$ is given by~\cite{rggr.conf.373.97}

\begin{equation}
\mathrm{SNR} = \sqrt{\frac{Y^2}{\sigma^2_Y}} =\frac{3 H_0^2}{10 \pi^2} \sqrt{T_{\mathrm{obs}}} \left[\int_{f_{\min}}^{f_{\max}} \frac{\gamma^2(|f|) \Omega_{\rm gw}^2(|f|)}{f^6 P_1(|f|)P_2(|f|)} \dif f \right]^{1/2} ,
\label{eq:ccSNR}
\end{equation}

\noindent where we usually assume a filter of the form

\begin{equation}
\Omega_{\mathrm{gw}}(f) = \Omega_{\alpha} (f/f_{\mathrm{ref}})^{\alpha}.
\label{eq:omegaTemplate}
\end{equation}

For this MDSC, we set $f_\mathrm{min}$ = 10Hz, $f_\mathrm{max}$ = 250Hz, use a reference frequency of $f_\mathrm{ref}$ = 100Hz and set $\alpha = 2/3$ as this is the theoretical value produced from the inspiral phase of CBCs.
In the right hand plot of Fig.~\ref {fig:background} we show the fractional SNR build-up as a function of frequency for different detector pairs.
We see that for all three pairs we reach to nearly 100\% of the total SNR by 120Hz, which is contained well within the limits we have set above for the analysis.


\section{Parameter Estimation}
\label{sec:PE}

Parameter estimation of signal models requires, at first, GW detection with high significance.
In this analysis, we use a method for parameter estimation of a SGWB background.
We seek to address the question of how well we can fit the model parameters.
As a concrete example, we show the recovery of parameters from the MDSC injection sets using both a power-law, $\alpha$, and CBC model.
We show how to estimate parameters such as a SGWB amplitude and the CBC coalescence rate.
To do so, we use a method presented in~\cite{prl.109.171102.12}, that introduced a maximum likelihood technique to simultaneously estimate multiple parameters of SGWB models using CC data from pairs of GW detectors.
This technique was used on recent results from LIGO to produce the first simultaneous limits on multiple parameters for power-law and CBC models of the SGWB, and to estimate the sensitivity of second-generation GW detectors to these models.

The likelihood function is defined as

\begin{equation}
L(\hat{Y}_i,\hat{\sigma}|\vec{\theta}) \propto \left[-\frac{1}{2} \sum_i \frac{\left(\hat{Y}_i - \Omega_M(f_i;\vec{\theta})\right)^2}{\sigma_i^2} \right] ,
\label{eq:Likelihood}
\end{equation}

\noindent where $\Omega_M(f_i;\vec{\theta})$ is the template spectrum that we are trying to fit by varying the parameters $\theta$, the sum runs over frequency bins $f_i$, which we set to be 0.25Hz, and $\hat{Y}_i$ and $\sigma_i^2$ are the estimator and variance in the frequency bin $i$.
The two methods, one a stochastic template-based analysis and the other a CC analysis, are very similar.
Traditional SGWB cross-correlation searches have adopted a particular power-law model, assuming a specific spectral index value and searching over the spectral amplitude.
The template-based maximum-likelihood estimator instead generically incorporates any model for the purposes of both detection and parameter estimation.
Therefore, it may be particularly useful for compact binary backgrounds.

\subsection{SGWB Models}

The amplitude and the frequency dependence of the SGWB spectrum depend on the physics of the model that generated it.
For example, in the CBC model, the spectrum is determined by the coalescence rate of binary systems throughout the universe and by the distribution of their chirp masses.
Past SGWB searches, performed using data from the initial LIGO and Virgo detectors~\cite{nat.460.990.09,prd.85.122001.12,prl.89.231101.14}, assumed a power-law model (see Eq.~(\ref{eq:omegaTemplate})) and set limits only on the amplitude $\Omega_{\mathrm{ref}}$ for fixed values of the spectral index $\alpha$ and of the reference frequency $f_\mathrm{ref}$.
This is reasonable as most SGWB models predict a power-law dependence in the LIGO-Virgo frequency band.

As discussed above, compact binary coalescences are among the most promising sources of gravitational waves for ground-based gravitational-wave detectors.
While detections of individual compact binaries are possible, another possibility is the detection of contributions from all CBCs in the universe to a SGWB.
The model we use is the average version of Eq.~(\ref{eq:omega_discrete}) where the discrete sum over the sources is replaced by an integral over the redshift, masses, sky position, inclination angle and $\psi$.
It was shown in~\cite{prl.109.171102.12} that it was sufficient to use only the average chirp mass $\mathcal{M}$ to determine the spectrum.
The model then consists of $\mathcal{M}$ and $\lambda$, which we recall is the mass fraction parameter, proportional to the local CBC rate per unit volume.


\section{Results}
\label{sec:results}

We now present the results from our analyses of the various mock data sets.
We first discuss the results from the analysis of the Gaussian noise data (data set 0), which we consider to be an ideal case.
We compare these against the results from the re-coloured noise which can be considered as a more realistic case.
We then detail the results from the simulation of the observing scenario.
Finally we show the results from the parameter estimation.

The results of the Gaussian and re-coloured data sets, where we use the same set of injections for both, are reported in Table~\ref{tab:Results}.
The first column lists the three detector pairs as well as the combined results which is the weighted sum of the three pairs where the combined point estimate is calculated using

\begin{equation}
Y_\mathrm{combined} = \frac{\sum\limits_{AB} Y_{AB}\sigma^{-2}_{AB}}{\sum\limits_{AB} \sigma^{-2}_{AB}},
\end{equation}

\noindent where $AB$ run over the three possible detector pairs and the combined error is given by

\begin{equation}
\sigma^{-2}_\mathrm{combined} = \sum_{AB} \sigma^{-2}_{AB}.
\end{equation}

\noindent The second, third and fourth columns give the estimated GW energy density, the error on this estimate and the SNR of the measurement for the data sets using Gaussian noise.
The fifth, six and seventh columns give the corresponding results for the data sets using re-coloured noise.
We define an SGWB signal as being detectable once it passes an SNR threshold value of 3, as defined in Eq.~(\ref{eq:ccSNR}).

\begin{table*}
\caption{\label{tab:Results}Results from all mock data sets.
The first column indicates the detector pair used in the analysis.
The second column gives the estimated value of $ \Omega_{\alpha}$ from the Gaussian data sets.
The third column gives the error on the measurement from the Gaussian data sets.
The fourth column gives the SNR of the detection from the Gaussian data sets.
The fifth column gives the estimated value of $ \Omega_{\alpha}$ from the re-coloured data sets.
The sixth column gives the error on the measurement from the re-coloured data sets. These are all nearly identical across the data sets because the contributions of signal to the overall noise background are minimal (of order $0.1\%$ when compared
to the instrumental noise). This was verified by performing the same analysis on a small subset of the data but containing only signal. 
The final column gives the SNR of the detection from the re-coloured data sets.
The table is divided by horizontal rows for the various data set with the injected value of $ \Omega_{\alpha}$, as calculated by Eq.~(\ref{eq:omega_discrete}), also given.}
\begin{center}
\begin{tabular}{ c | c c c | c c c}
\hline \hline
Detector Pair & Point Estimate & Error & SNR & Point Estimate & Error & SNR\\ \hline
 &  \multicolumn{3}{c|}{Gaussian } & \multicolumn{3}{c}{Re-coloured Noise} \\ \hline
\multicolumn{7}{c}{Set 0 (Noise only): $ \Omega_{\alpha} = 0$} \\ \hline
HL            & $-8.093\times10^{-10}$ & $1.473\times10^{-9}$ & -0.55 & $-1.119\times10^{-9}$   & $1.683\times10^{-9}$ & -0.66   \\
HV            & $-1.04\times10^{-8}$     & $1.139\times10^{-8}$ & -0.91 & $-1.12\times10^{-8}$ & $1.407\times10^{-8}$ & -0.8 \\
LV             & $-1.47\times10^{-9}$     & $1.042\times10^{-8}$ & -0.14 & $-1.765\times10^{-9}$ & $1.38\times10^{-8}$ & -0.13   \\ \hline
Combined & $-9.769\times10^{-10}$ & $1.447\times10^{-9}$ & -0.68 & $-1.268\times10^{-9}$   & $1.659\times10^{-9}$   & -0.76   \\ \hline
\multicolumn{7}{c}{Set 1 (BNS): $ \Omega_{\alpha} = 1.364\times10^{-8}$} \\ \hline
HL            & $1.512\times10^{-8}$ & $1.474\times10^{-9}$ & 10.26 & $1.455\times10^{-8}$   & $1.683\times10^{-9}$ & 8.65 \\ 
HV            & $7.706\times10^{-9}$ & $1.139\times10^{-8}$ & 0.68   & $-9.858\times10^{-9}$ & $1.235\times10^{-8}$ & -0.8 \\ 
LV             & $5.491\times10^{-9}$ & $1.042\times10^{-8}$ & 0.53   & $7.451\times10^{-9}$   & $1.235\times10^{-8}$ & 0.6 \\ \hline
Combined & $1.481\times10^{-8}$ & $1.448\times10^{-9}$ & 10.23 & $1.399\times10^{-8}$   & $1.653\times10^{-9}$ & 8.46 \\ \hline
\multicolumn{7}{c}{Set 2 (BNS): $ \Omega_{\alpha} = 1.411\times10^{-8}$} \\ \hline
HL            & $1.573\times10^{-8}$ & $1.474\times10^{-9}$ & 10.68 & $1.601\times10^{-8}$ & $1.69\times10^{-9}$ & 9.47 \\
HV            & $7.713\times10^{-9}$ & $1.139\times10^{-8}$ & 0.68   & $-6.738\times10^{-9}$ & $1.24\times10^{-8}$ & -0.54 \\
LV             & $5.854\times10^{-9}$ & $1.042\times10^{-8}$ & 0.56   & $1.131\times10^{-9}$ & $1.24\times10^{-8}$ & 0.09 \\ \hline
Combined & $1.541\times10^{-8}$ & $1.447\times10^{-9}$ & 10.65 & $1.533\times10^{-8}$ & $1.659\times10^{-9}$ & 9.24 \\ \hline
\multicolumn{7}{c}{Set 3 (BBH): $ \Omega_{\alpha} = 6.975\times10^{-9}$} \\ \hline
HL            & $5.175\times10^{-9}$ & $1.474\times10^{-9}$ & 3.51 & $4.725\times10^{-9}$   & $1.683\times10^{-9}$ & 2.81 \\
HV            & $4.257\times10^{-9}$ & $1.139\times10^{-8}$ & 0.37 & $-6.117\times10^{-9}$ & $1.407\times10^{-8}$ & -0.43 \\
LV             & $1.763\times10^{-9}$ & $1.042\times10^{-8}$ & 0.17 & $3.968\times10^{-9}$   & $1.379\times10^{-8}$ & -0.29 \\ \hline
Combined & $5.094\times10^{-9}$ & $1.447\times10^{-9}$ & 3.52 & $4.448\times10^{-9}$   & $1.659\times10^{-9}$ & 2.68 \\ \hline
\multicolumn{7}{c}{Set 4: (BNS+BBH): $ \Omega_{\alpha} = 2.022\times10^{-8}$} \\ \hline
HL            & $2.056\times10^{-8}$ & $1.474\times10^{-9}$ & 13.94 & $1.991\times10^{-8}$   & $1.684\times10^{-9}$ & 11.83 \\
HV            & $9.674\times10^{-9}$ & $1.139\times10^{-8}$ & 0.85  & $6.452\times10^{-9}$  & $1.352\times10^{-8}$ & 0.48 \\
LV             & $7.211\times10^{-9}$ & $1.042\times10^{-8}$ & 0.69  & $1.777\times10^{-8}$  & $1.328\times10^{-8}$ & 1.34 \\ \hline
Combined & $2.012\times10^{-8}$ & $1.448\times10^{-9}$ & 13.9  & $1.968\times10^{-8}$  & $1.658\times10^{-9}$ & 11.87 \\ \hline
\multicolumn{7}{c}{Observing scenario: $ \Omega_{\alpha} = 2.756\times10^{-9}$} \\ \hline
HL            & $3.581\times10^{-9}$ & $6.869\times10^{-10}$ & 5.21 & -------- & -------- & --- \\
HV            & $9.413\times10^{-10}$ & $6.207\times10^{-9}$ & 0.15 & -------- & -------- & --- \\
LV             & $4.235\times10^{-10}$ & $ 5.723\times10^{-9}$ & 0.07 & -------- & -------- & --- \\ \hline
Combined & $3.505\times10^{-9}$ & $6.779\times10^{-10}$ & 5.17 & -------- & -------- & --- \\
\hline \hline
\end{tabular}
\end{center}
\end{table*}


\subsection{Gaussian}

The results from the analysis of all the Gaussian data sets are presented in the left-hand section of Table~\ref{tab:Results}.
The first results we highlight are from data set 0, which is our control set as it consists of just independent  Gaussian noise.
The measurement of $ \Omega_{\alpha}$ for the three detector pairs and the combined result give very low estimates and are the results of statistical fluctuations in the Gaussian data.
The main result from this data set is the measurement of the error which we note is consistent with the error measurements of each of the other Gaussian noise data sets.

Data sets 1 and 2, where we only consider a population of BNS with a high merger rate, were designed to test how much bias is added to the measurements of $ \Omega_{\alpha}$ when we neglect to remove the loud detectable signals from the data streams\footnote{In this MDSC we can choose to simply not include any individually detectable signals within the data streams.
In reality, removing a detected signals from the data streams is very difficult as there may be some inaccuracy in measuring its true parameter used to produce the waveform which would leave some residual signal.
Instead, we simply do not analyse the data that is known to contain the signal in the frequency band that we are searching.} compared to when we only use sub-threshold signals.
By keeping the detectable events, we also increase $ \Omega_{\alpha}$, as indicated in Table~\ref{tab:Results}.
We find that in both cases we are able to measure the background estimate to within 1$\sigma$ of the true value, as well as obtaining similar SNRs, both greater than 10.

With data set 3, where we consider a population of only BBH with a lower merger rate than before, we find that we are still able to make a detection but with a lower SNR of 3.52.
This also gives the largest error in the measurement of $ \Omega_{\alpha}$ with a measured value 1.3$\sigma$ away from the true value.
These results show that the estimation of $ \Omega_{\alpha}$ is still possible given a non-continuous GW signal (see middle plot of Fig.~\ref{fig:timeSeries}).

\begin{figure*}
\hskip -0.3cm
\includegraphics[width=0.45\textwidth]{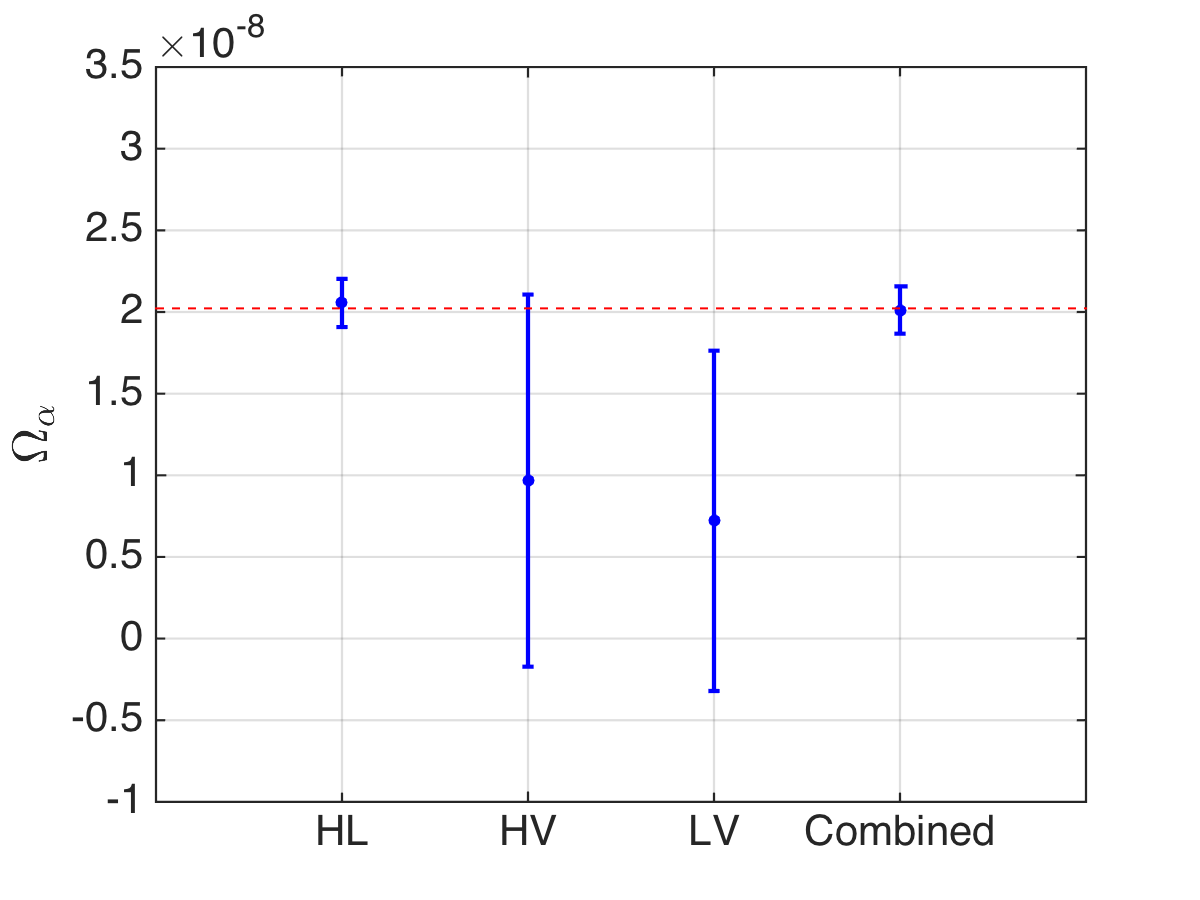}
\hskip -0.5cm
\includegraphics[width=0.45\textwidth]{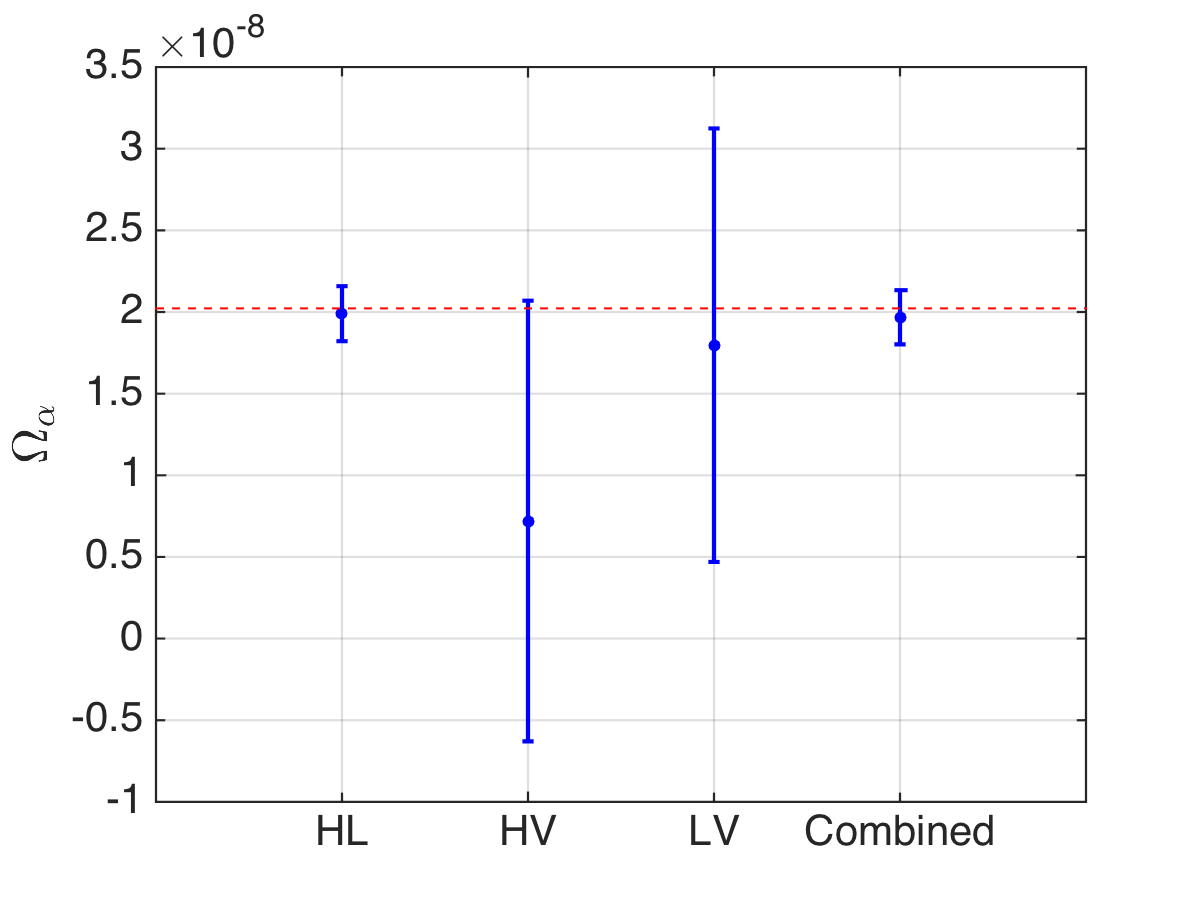}
\caption{\emph{Left}--- Results for the year-long data set 4 with Gaussian noise where we show the measured value of $ \Omega_{\alpha}$ with error bars included for each of the detector pairs as well as the combined result of the weighted sum of the three detector pairs.
The horizontal red dashed line shows the true injected value of $ \Omega_{\alpha} = 2.022\times10^{-8}$ for this data set.
\emph{Right}--- Results for the year-long set 4 with re-coloured noise where we show the measured value of $ \Omega_{\alpha}$ with error bars included for each of the detector pairs as well as the combined result of the weighted sum of the three detector pairs.
The horizontal red dashed line shows the true injected value of $ \Omega_{\alpha} = 2.022\times10^{-8}$ for this data set.}
\label{fig:GaussRes}
\end{figure*}

Data set 4, which is the combination of data set 1 and data set 3, gives a measured $ \Omega_{\alpha}$ spectrum equal to the sum of the results from data sets 1 and 3.
A plot of the results for the three detector pairs, as well as the combined results, are shown in the left-hand plot of Fig.~\ref{fig:GaussRes}, where the measured  values $ \Omega_{\alpha}$are plotted in blue along with their respective error bars.
Here the red dashed line shows the true value of $ \Omega_{\alpha}$.
We see here that we are able to recover the value of $ \Omega_{\alpha}$ to within 1.6\% with the HL detector pair and 0.5\% when we consider the combined results.
This is likely due to the very high SNR of 13.9.

\subsection{Re-coloured}

The re-coloured data consists of the re-coloured initial LIGO and Virgo detector noise combined with the same data sets signal as for the Gaussian case.
This analysis more closely simulates the likely output from the advanced detectors, which will suffer from various environmental noise sources contaminating the data.
The results from the analysis of all the re-coloured data sets are presented in the right-hand section of Table~\ref{tab:Results} (next to the Gaussian results for easy comparison).
As in the case of the Gaussian data set, set 0 contains only noise and serves as a baseline for the analysis.
It has a combined point estimate well within $1 \sigma$ of 0.
The results for data sets 1 and 2 are also consistent with the Gaussian sets.
The SNR for these sets are about 20\% lower than that of the Gaussian set, which is due to the non-Gaussian noise transients.
They are also within $1 \sigma$ of the true values.
Data set 3 and 4 show similar effects to data set 2, with a lower SNR than in the Gaussian case and with a recovered value of $ \Omega_{\alpha}$ within 1.6 $\sigma$ of the true values for data set 3 and within 1$\sigma$ for data set 4.


\subsection{Observing Scenario}

\begin{figure*}
\hskip -0.3cm
\includegraphics[width=0.45\textwidth]{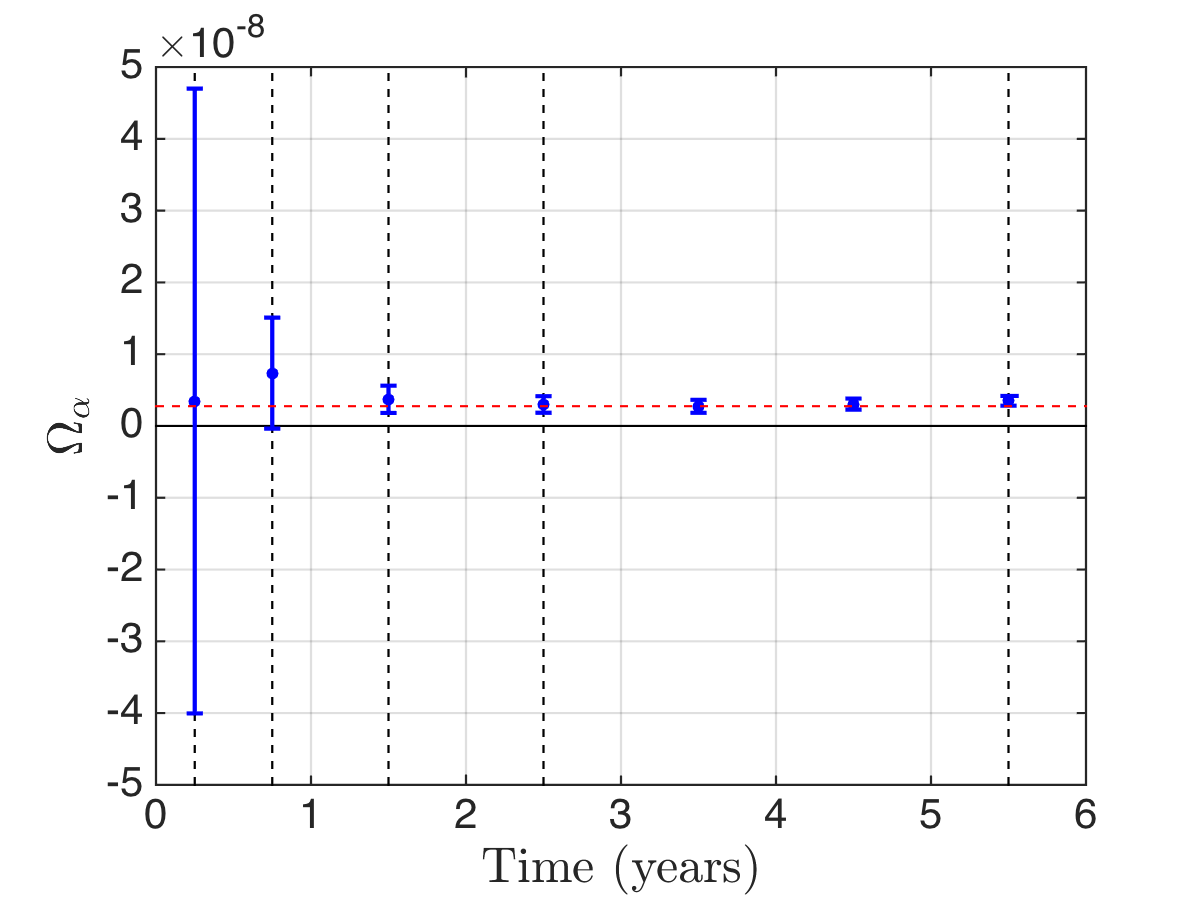}
\hskip -0.5cm
\includegraphics[width=0.45\textwidth]{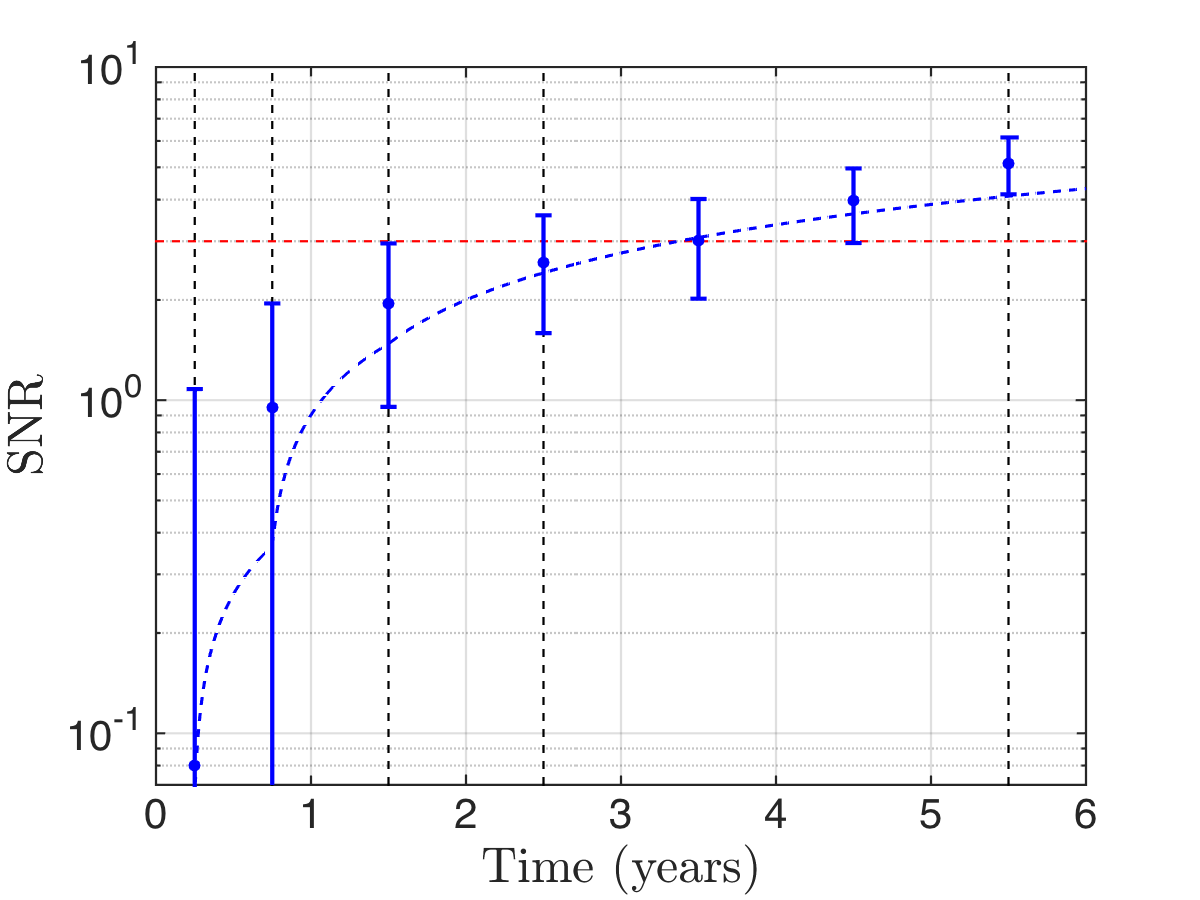}
\caption{\emph{Left}--- Combined results for the estimated value of $ \Omega_{\alpha}$ as a function of time for the 5.5 long year observing scenario.
The black dashed lines show the end point of each observing run, the red dashed line is the true injected value of $ \Omega_{\alpha} = 2.756\times10^{-9}$, the horizontal black solid line is $ \Omega_{\alpha}$ = 0 and the blue points are the measured $ \Omega_{\alpha}$ values with their error bars.
\emph{Right}--- Combined results for the measured SNR as a function of time for the 5.5 year long observing scenario.
The black dashed lines show the end point of each observing run, the red dashed line represents SNR = 3, which we use as a threshold for claiming a detection, the blue dashed line shows the theoretical SNR as a function of time given by Eq.~(\ref{eq:ccSNR}) and the blue points are the combined SNRs corresponding 1$\sigma$ error bars.}
\label{fig:OSResults}
\end{figure*}

We report the results from the observing scenario where we consider evolution of the detector sensitivities during different observational runs, the results of which are shown in Fig.~\ref{fig:OSResults}.
In the left-hand plot we show the measured value of $ \Omega_{\alpha}$ as a function of observation time with error bars included.
The black dashed lines represent the start/end of each phase as detailed in Section~\ref{sec:dataSets} and the red dashed line represents the true injected $ \Omega_{\alpha}$ value while the solid black horizontal line represents $ \Omega_{\alpha}$ = 0.
In the right hand plot we show the combined SNR as a function of time for the same points in the left-hand plot, as well as the theoretical SNR given by Eq.~(\ref{eq:ccSNR}) using the blue dashed line.
Here again the vertical black dashed lines show the start/end of each phase and the red dashed line indicates SNR = 3 which we use as the threshold value for claiming a detection of the SGWB.

The first result to note is that over the course of the whole 5.5 years of the observing run the size of the error bars reduces significantly.
By the end of the first 1.5 years of observations they have already reduced by over an order of magnitude.
The second result to note is that already after the first 1.5 years of observation, we will observe disagreement with the null result at 95\% confidence ($2\sigma)$.
The third result, which is in agreement with the theoretical model, is that we may be able to confirm the detection of the SGWB with an SNR $\geq$ 3 after a period of about 3.5 years.
At the end of the 5.5 year observing run, for this coalescence rate, we report that we have a total SNR of 5.15.
We note that, in the right-hand plot of Fig.~\ref{fig:OSResults}, the measured SNR (0.95) for the end of the second phase is well above what is predicted by the theoretical model (SNR = 0.35), although it is still within the $1\sigma$ range.
This is explained by the larger than average measurement of $ \Omega_{\alpha}$ at the end of the second phase, as shown in the left-hand plot of Fig.~\ref{fig:OSResults}, which, whilst being significantly larger than other measurements, is still within $1\sigma$ of the true value.

\subsection{Parameter Estimation}
\label{sec:peresults}

\begin{figure*}
\hskip -0.3cm
\includegraphics[width=0.45\textwidth]{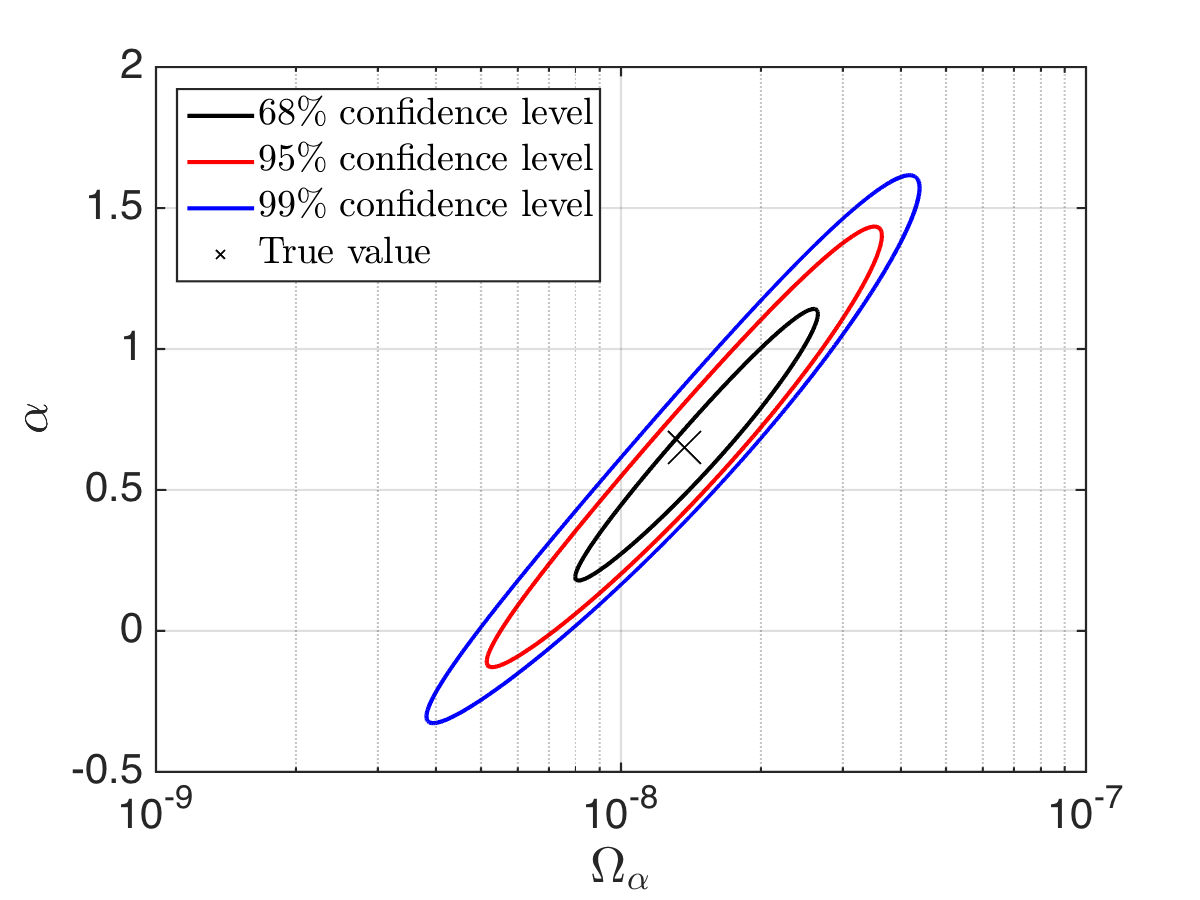}
\hskip -0.5cm
\includegraphics[width=0.45\textwidth]{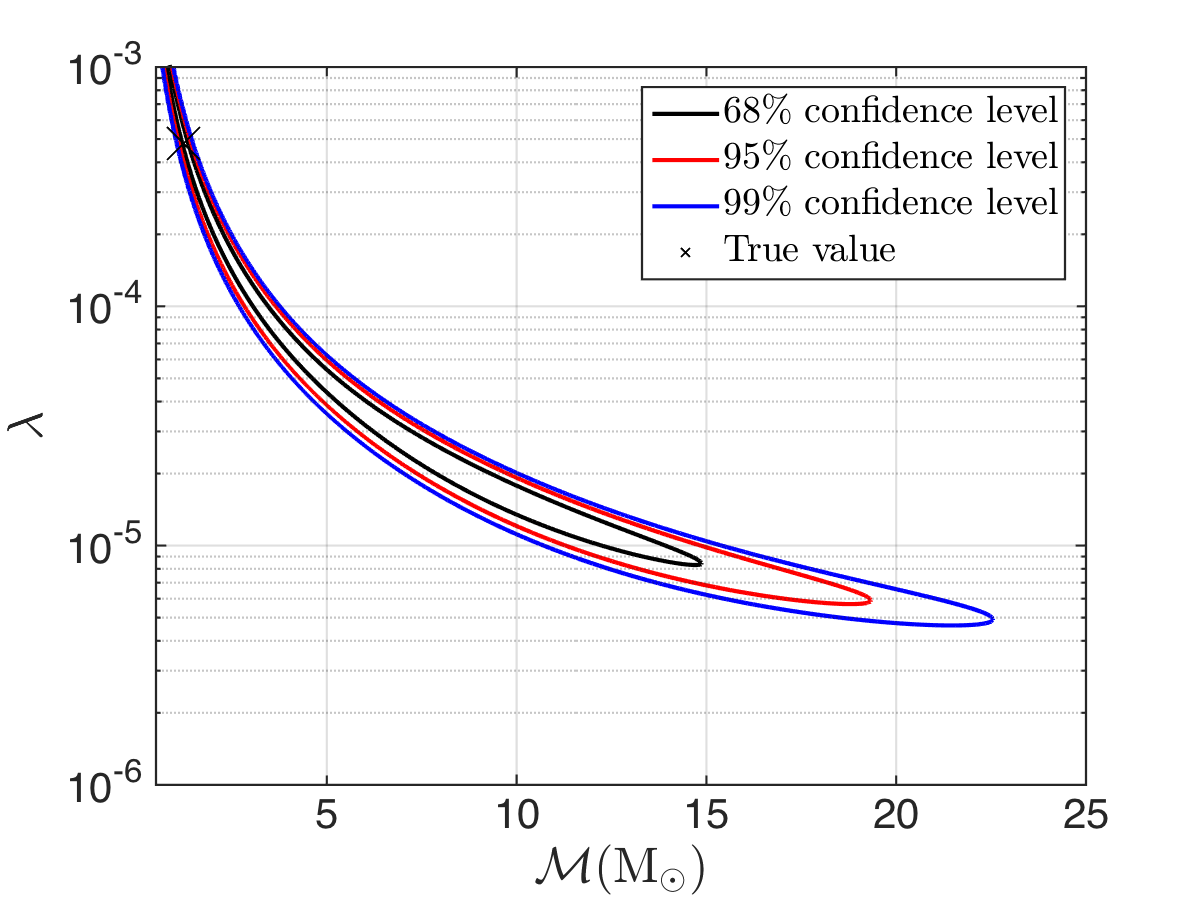}
\caption{\emph{Left}--- The posteriors for amplitude, $ \Omega_{\alpha}$, and its spectral index, $\alpha$, for data set 1 with the 99\% confidence level (blue), 95\% confidence level (red) and 68\% confidence level (black) shown.
We show the true parameter values with an ``X".
\emph{Right}--- The posteriors for average chirp mass, $\mathcal{M}$, and coalescence rate, $\lambda$, for data set 1 with the 99\% confidence level (blue), 95\% confidence level (red) and 68\% confidence level (black) shown.
We denote the correct CBC parameters by an ``X''.}
\label{fig:Posteriors}
\end{figure*}

In order to construct parameter posterior distributions, we use models for $\Omega_M(f_i;\vec{\theta})$ for various sets of waveform parameters.
We use a power-law and compact binary model for $\Omega_M(f_i;\vec{\theta})$.
Eq.~(\ref{eq:Likelihood}) is evaluated repeatedly for each set of parameters and is maximized for those parameters that best fit the data.
Parameter posterior distributions are constructed for parameter sets of equal likelihood.
Example posteriors are shown in Fig.~\ref{fig:Posteriors} for the power-law and CBC models.
Table~\ref{tab:PETable} shows results for all of the injection sets.
We provide parameter estimates for the power-law model where, for CBC systems, the power-law index is $\alpha=2/3$.
We also provide parameter estimates and the true values for the CBC model.
The constraints on the CBC background are relatively weak and highly dependent on the mass of the system (see the right hand plot of Fig.~\ref{fig:Posteriors}).
Therefore, the limits we place on chirp mass and coalescence rate are in terms of bounds on the parameters.
The bounds we place are consistent with the injected parameter values.
We tested the case where we consider multiple CBC models, as was used to produce data set 4, and found that the posteriors are broadened by a significant amount.

\begin{table*}
\caption{\label{tab:PETable} Parameter estimation results for the various data sets.
We provide the 99\% confidence limits for both the power-law and CBC models, as well as the injected parameters.
The first column indicates the data set.
The second column is the estimated amplitude of $ \Omega_{\alpha}$.
The third column is the estimated power-law of the signal.
The fourth and fifth columns gives the injected values of the amplitude and power-law.
The sixth column give the estimated average chirp mass.
The seventh column gives the estimated rate of events.
The eighth and ninth columns gives the injected values of the average chirp mass and rate of events.}
\begin{center}
\begin{tabular}{ c c c c c c c c c}
\hline \hline
Data Set & $ \Omega_{\alpha}$ & $\alpha$ & True $ \Omega_{\alpha}$ & True $\alpha$ & $\mathcal{M}$ & $\lambda$   & True $\mathcal{M}$ & True $\lambda$ \\ \hline
Set 1 & [$4.0 \times 10^{-10}$, $4.0 \times 10^{-8}$] & [-0.3, 1.6] & $1.5 \times 10^{-8}$ & 0.65 & $\leq 23$   & $\geq 1.0$   & 1.22 & 10  \\
Set 2 & [$4.0 \times 10^{-9}$,   $4.0 \times 10^{-8}$] & [-0.3, 1.6] & $1.5 \times 10^{-8}$ & 0.65 & $\leq 22$   & $\geq 0.9$   & 1.22 & 10 \\
Set 3 & [$1.0 \times 10^{-12}$, $3.4 \times 10^{-8}$] & [-4, 2]       & $7.6 \times 10^{-9}$ & 0.73 & $\leq 100$ & $\geq 0.04$ & 8.7   & 0.3 \\ 
Set 4 & [$8.7 \times 10^{-9}$,   $4.8 \times 10^{-8}$] & [0, 1.4]      & $2.1 \times 10^{-8}$ & 0.68 & $\leq 19$   & $\geq 1.6$   & 1.37 & 10.2 \\
\hline \hline
\end{tabular}
\end{center}
\end{table*}

\section{Conclusion}
\label{sec:conclusion}

In this SGWB MDSC for the advanced detectors, LIGO and Virgo, we have presented our methods for the productions and analysis of multiple mock data sets, as well as the results and their scientific interpretation.
We consistently find that the best results are obtained with the aLIGO detector pair (HL), compared to the aLIGO-AdVirgo pairs (HV, LV), though we still see some slight improvement when we consider the combined results from all detector pairs.
This is in agreement with what we expect given the difference in sensitivities and orientations of the pairs.
In the case of our three detector network, the two aLIGO detectors are the best aligned and have the smallest separation.

We have made comparisons between the use of Gaussian detector noise, which can be considered an idealistic case, and re-coloured noise data which is expected to be more realistic.
We injected the same sources into both sets of noise to ensure that we are measuring the same signal.
In both cases, we have been able to recover the injected  value $ \Omega_{\alpha}$ to within 2$\sigma$ using each data set.

From the analysis of the observing scenario data set we have shown that for the optimistic values of the CBC event rate the first deviation from the null result (at 95\% confidence) may be observed as early as 1.5 years into the observation time.
This assumes that we are able to achieve the designed sensitivities at the end of each observing phase, as outlined in Section~\ref{sec:dataSets}, and that the coalescence rate of CBC signals is significantly large as to make it detectable within a few years of operations at design sensitivity.

The results from data sets 1 and 3 have also shown that the theoretical prediction given in~\cite{prd.89.084063.14} holds true when applied to mock data.
That is, the the statistical properties of the CBC GW signals, whether it be a continuous signal or more popcorn like, do not matter when we make a measurement of $\Omega_{\alpha}$; all that is important is the total number of events that coalesce within the observational period and the GW energy spectrum emitted by each event.

Finally, we have shown that we are able to apply parameter estimation methods to the data in order to place confidence levels on different parameters.
The detection of a stochastic signal will not be able to provide enough information by itself to place tight constraints on these parameters, but when considered in combination with detections of single events, it can become a very useful tool to explore the ensemble of sources from the whole universe~\cite{prl.109.171102.12}.

The results from the estimation of the average chirp mass and coalescence rate also show that there is equal probability of having a high rate of events and low average mass as having a low rate of events and high average mass.
In both cases the amplitude of the signal and and the spectral index will be the same for the frequency range we search over, but in the first case the GW signals will be continuous while in the second case the GW signals will be highly non-Gaussian and popcorn like.
The isotropic CC search we implement here is insensitive to two types of signals as it just considers the average strength of the signal over the observational period.
In order to be able to differentiate these two signal types a non-Gaussian analysis must be developed that is able to search over both time and frequency~\cite{prd.87.043009.13}.

Future MDSCs may wish to explore several areas that have not been covered here, such as, the inclusion of intermediate mass black holes (IMBH), which may coalesced in the middle of the frequency search band, and therefore, given a high enough rate, may affect the analysis.
Or, we could add a loud SGWB signal of cosmological origin that has a spectral index that differs from that of the astrophysical contribution.
Another important question would be to investigate the behaviour of the CC analysis when correlated noise between different pairs of detectors is included in the mock data~\cite{prd.87.123009.13,prd.90.023013.14} .
The continuation of MDSCs will be an important part of the verification process for LIGO and Virgo when a SGWB is eventually observed.


\begin{acknowledgments}
DM acknowledges the PhD financial support from the Observatoire de la C$\rm{\hat{o}}$te d’Azur and the PACA region.
MC was supported by the National Science Foundation Graduate Research Fellowship Program, under NSF grant number DGE 1144152.
SM and JDR were supported NSF grants CREST HRD-1242090 and PHY-1205585.
JDR also acknowledges support from NSF grants PHY-1505861 and
PHY-1066293, and the hospitality of the Aspen Center for Physics.
NLC is supported by NSF grants PHY-1505373 and PHY-1204371.
VM is supported in part by NSF grant PHY-1204944 at the University of Minnesota.
The stochastic GW search has been carried out using the MatApps software available at \url{https://www.lsc-group.phys.uwm.edu/daswg/projects/matapps.html}.
\end{acknowledgments}


\bibliography{bibliography}

\end{document}